\documentclass[pre,floatfix,twocolumn,superscriptaddress]{revtex4-1}

\usepackage{latexsym}
\usepackage{amsmath, amsthm, amssymb}
\usepackage{mathrsfs}
\usepackage{epsfig}
\usepackage{graphicx}
\usepackage{dcolumn}
\usepackage{expl3}
\usepackage{float}

\usepackage{hyperref}
\usepackage{placeins}

\allowdisplaybreaks

\begin{document}

\title{Uncovering the invariant structural organization of the human connectome}

\author{Anand Pathak}
\affiliation{The Institute of Mathematical Sciences, CIT Campus, Taramani, Chennai 600113, India}
\affiliation{Homi Bhabha National Institute, Anushaktinagar, Mumbai 400 094, India}

\author{Shakti~N.~Menon}
\affiliation{The Institute of Mathematical Sciences, CIT Campus, Taramani, Chennai 600113, India}

\author{Sitabhra~Sinha}
\affiliation{The Institute of Mathematical Sciences, CIT Campus, Taramani, Chennai 600113, India}
\affiliation{Homi Bhabha National Institute, Anushaktinagar, Mumbai 400 094, India}

\date{\today}


\begin{abstract} 
In order to understand the complex cognitive functions of the human brain, it is essential to 
study the structural macro-connectome, i.e., the wiring of different brain regions to each other through axonal pathways, that has been revealed by
imaging techniques.
However, the high degree of plasticity 
and cross-population variability in human brains makes it difficult to relate structure to function, motivating a search
for invariant patterns in the connectivity. At the same time,
variability within a population can provide
information about the generative mechanisms. 
In this paper we analyze the connection topology and 
link-weight distribution of human structural connectomes
obtained from a database comprising $196$ subjects. 
By demonstrating a correspondence between the occurrence frequency of individual links
and their average weight across the population, we show that the process by which the 
human brain is wired is not 
independent of the process by which the link weights of the connectome are determined. 
Furthermore, using the specific distribution of the 
weights associated with each link over the entire population,
we show that a single parameter that is specific to a link can account for its frequency of occurrence, as well as, the variation in its weight across different subjects. 
This parameter provides a basis for ``rescaling'' the link weights in each connectome,
allowing us to obtain a generic network representative of the 
human brain, distinct from a simple average over the connectomes.
We obtain the functional connectomes by
implementing a neural mass model on each of the vertices of the corresponding
structural connectomes. 
By comparing these with the empirical functional brain networks, 
we demonstrate that the rescaling procedure yields a closer
structure-function correspondence. Finally, we show that the representative network 
can be decomposed into a \textit{basal} component that is stable across
the population and a highly variable \textit{superstructure}.
\end{abstract}

\maketitle


\section{Introduction}
One of the key goals of neuroscience from its very inception has been
to unravel the workings of the human brain. Not only is this 
of great scientific interest and significant from a philosophical perspective,
but it is important also for informing clinical and psychiatric practice. 
Studying the nervous 
systems of non-human model organisms do allow us to gain an understanding of
fundamental aspects of their development, structure and function.
Moreover, this has provided us with numerous insights on how a 
system as complex as the brain could have evolved, and the associated emergence
of behavior such as cognition. However, there are
limitations to how phenomena observed in relatively simpler nervous systems
can be generalized to those with much higher complexity. For instance
the ocular dominance columns of the visual cortex that occur in monkey
and cat brains are not seen in mice or rat brains.~\cite{Horton2005}.
Similarly, the anatomical and functional organization of a macaque brain is
vastly different from that of a human brain~\cite{Passingham2009}. In fact
there are several fundamental aspects of human brain structure and function
that are unique to the species~\cite{Rilling2014, Kaas2013}. Therefore, in 
spite of the ethical and technological bottlenecks that hinder the study of
the human brain to the level of detail and precision as can be achieved in
other mammals and invertebrates, it is crucial that techniques for 
analyzing and interpreting the structure and function of human brains
at multiple length and time scales continue to be developed and refined.

One of the primary approaches that is commonly used to study a brain
is to describe its macro-scale connectome~\cite{Sporns2005}, i.e., the 
structure of connectivity between distinct brain regions through
axonal tracts. Several studies have pointed to the essential role of
such a ``wiring diagram'' of the nervous system as a foundational model
in understanding functional localization at multiple levels, ranging
from molecular and cellular up to systems and behavioral 
levels~\cite{Swanson2010}. As shown in~\cite{Pathak2020a},
the wiring diagrams of neuronal connectivity can provide insights on
the processes that underlie brain development, as well as, the 
functional implications arising from their structural organization~\cite{Pathak2020}. 
However, one of the characteristic features of the human structural 
connectome is its large variability across   
individuals~\cite{Sporns2005, Llera2019, Smith2019} and over 
time~\cite{Madole2020}. This variability in the structural
connectome could be a key factor in understanding the generative mechanism 
underlying the development of brain structure~\cite{Klimm2014}. The 
diversity in structural connectivity is significantly smaller than that
of functional connectivity~\cite{Kelly2012, Yan2013}, which is determined 
by temporal correlations between the electrophysiological activity of different
brain regions. These variabilities are important in studying the 
relationship between structure and function in human brains, as not only
is structural connectivity known to affect brain
function~\cite{Honey2009, Hagmann2008}, but function has been shown to 
influence the structure as well~\cite{Mishra2019, Taya2015}. However, despite the
high variability of brain networks in humans, it has been observed that 
certain structural features are universal~\cite{Bullmore2009}. Thus, it
is pertinent to ask whether we can describe a typical ``representative''
structural connectome for a human brain. Such a representative network 
might not only be useful in studying fundamental aspects of the 
structure-function relationship in the human brain, but also the deviations
from this ``basic plan'' in a connectome might reveal structural correlates of 
functional impairments leading to clinical disorders.

In this work, we study a large ensemble of brain connectivity networks
that were obtained from a cohort of $196$ healthy human subjects through 
diffusion tensor imaging for the purpose of characterizing the variability 
in the structural connectome. We find that there is a
correspondence between the diversity of topological connectivities and the variation in
the distribution of connection weights, suggesting that the generative 
mechanisms giving rise to the ``wiring'' and those determining the weights are related. 
We further find that the connection strengths of links that 
are frequently found in the population are described by link-specific 
Poisson processes, indicating that the generative mechanism of a 
significant portion of the brain connectivity might involve  
independent discrete random processes. This allows us to reassign the link 
weights of structural connectivity, which would represent the discrete 
Poisson variables instead of original weights and thus obtain the rescaled 
weight matrices. Using the corresponding resting state functional 
connectivities obtained from the same cohort, we show that the structural 
connectome with rescaled weights consistently show better correspondence 
with the functional connectivity, suggesting that the rescaling process
might provide a more informative framework for interpreting the structural connectivity 
in terms of function. This also provides us with a means for determining the  
generic ``representative'' network describing a human connectome. Finally, 
we show that the representative network is intrinsically resolved into two 
components, one which is invariant across the population and another that 
exhibits a much higher degree of variability across individuals.

\section{Materials and Methods}

\subsection{Connectivity Data}
The human brain structural and functional connectivity dataset analyzed here 
has been derived from the {\em Nathan Kline Institute (NKI) / Rockland
Sample}~\cite{Nooner2012} -  a publicly available repository of 
diffusion tensor imaging (DTI) and resting state functional magnetic
resonance imaging (rs-fMRI) data - which was further processed into 
connectivity matrices and made publicly available in the {\em UCLA 
multimodal connectivity database} at
\url{http://umcd.humanconnectomeproject.org/}~\cite{Brown2012}.

The data comprises structural connectivity matrices $W$ and functional 
connectivity matrices $C$ obtained from $196$ healthy human subjects: $114$
male and $82$ female, with ages ranging from $4$ to $85$ years. Each matrix
describes a network comprising $188$ nodes that represent 188 brain regions
defined by parcellation of the entire gray matter region of the human brain 
(cerebral cortex, sub-cortical areas, cerebellum, brain stem etc.) using an
fMRI based clustering method~\cite{Craddock2012}. It also contains the
$3$-dimensional coordinates for each of the brain regions in a standardized
space. For structural connectivity (SC) matrices $W$, the connection
strengths $W_{ij}$ corresponding to the weighted undirected links $(i,j)$ 
represent the density of axonal bundles between brain regions $i$ and $j$ 
as obtained from DTI, while the connection strengths $C_{ij}$ in functional
connectivity matrices $C$ represent the Pearson's correlation coefficients between the 
time-series of dynamical activities in regions $i$ and $j$, as measured 
through blood oxygen level dependent (BOLD) imaging using fMRI.

\subsection{Rescaling to obtain Poisson distributed link weights}

A Poisson distribution of mean $\lambda$ for a random discrete variable
$X$ is given by:
\begin{equation}
P(X=k)  =  \frac{\lambda^k e^{-\lambda}}{k!}\,,
\end{equation}
\noindent
where the parameter characterizing the distribution $\lambda = \langle X \rangle = Var(X)$. 
Here the link weights 
$W_{ij}$ for a link between a pair of regions $(i,j)$ are considered
to be obtained by rescaling Poisson distributed variables $\mathcal{W}_{ij}$
that have mean $\lambda_{ij}$, as $W_{ij} = s_{ij}\mathcal{W}_{ij}$.
Such a rescaled Poisson variable has also been described 
in~\cite{Prem2017}. The mean and variance of the rescaled Poisson variables 
$W_{ij}$ are given by:
\begin{equation}
\begin{split}
\langle W_{ij} \rangle &= s_{ij} \lambda_{ij}\,,  \\
Var(W_{ij}) &= s^2_{ij} \lambda_{ij}\,,
\end{split}
\label{eq_sc}
\end{equation}
For a given distribution of weights $W_{ij}$ of a link $(i,j)$,  we 
determine $s_{ij}$ from Eq.~(\ref{eq_sc}) 
\begin{equation}
 s_{ij} = \frac{Var(W_{ij})}{\langle W_{ij} \rangle}
\end{equation}
Upon obtaining the rescaling factor $s_{ij}$ for a link $(i,j)$, we 
rescale the weights $W_{ij}$ across the population to obtain the Poisson 
distributed rescaled weights:
\begin{equation}
\mathcal{W}_{ij} = \lfloor \frac{W_{ij}}{s_{ij}} + 0.5 \rfloor
\end{equation}
where $\lfloor x+0.5 \rfloor$ gives nearest integer of $x$. The rescaled
weights are related to the Poisson parameter by
$\langle \mathcal{W}_{ij} \rangle = \lambda_{ij}$.

\subsection{Goodness of fit}
Assuming a rescaled Poisson distribution for all links $(i,j)$, we
initially calculate the Poisson parameter $\lambda_{ij}$, rescaling factor 
$s_{ij}$ and the rescaled weights $\mathcal{W}_{ij}$ for each link. We then use a
Pearson's Chi-squared test~\cite{Pearson1900} to determine whether
the rescaled weights obtained using the method described above fits a
theoretically expected Poisson distribution $\mathcal{P}(\lambda_{ij})$ for the corresponding 
$\lambda_{ij}$ with significant likelihood. First, we calculate the test 
statistic $\chi^2_{ij}$ for the rescaled weight frequency distribution of 
each link:
\begin{equation}
 \chi^2_{ij} = \sum_{k=1}^n \frac{(O_k - E_k)^2}{E_k}\,,
\end{equation}
where $n$ is total number of bins, $O_k$ is number of observations having 
$\mathcal{W}_{ij}=k$ and $E_{k}$ is the theoretically expected number of 
observations, assuming a Poisson distribution for $\mathcal{W}_{ij}$. 
Here, $E_k$ is obtained from $P(\mathcal{W}_{ij}=k)$ as:
$E_k = \lfloor N*P(\mathcal{W}_{ij}=k) +0.5\rfloor$, 
where $N$ is the total number of connectomes analyzed. If a bin has $E_k<5$, it is
merged with the adjacent bins, thus reducing the total number of bins. To
ensure the validity of the test, we require that the final number of bins 
$n\geqslant 3$. If the final number of bins is less than $3$, which may 
arise in the case of links with very low $\lambda_{ij}$, we consider the
link to be too rare for this statistical test. When the number of bins are 
sufficient, we compare the statistic $\chi^2_{ij}$ with the Chi-squared 
critical values for the upper tail one-sided test with significance value 
$\alpha=0.01$, obtained from
\url{https://www.itl.nist.gov/div898/handbook/eda/section3/eda3674.htm}.
Links having $\chi^2_{ij}$ values 
less than the corresponding critical values, one cannot reject the possibility that
they are from a Poisson distribution. 

\subsection{Partial fitting by excluding outlier data from deviating links}

For each link $(i,j)$ that deviated from a Poisson distribution, we
performed an iterative process where at each step the data point with the 
largest value of $\mathcal{W}_{ij}$ was removed, new values of
$\lambda_{ij}$, $s_{ij}$ and $\mathcal{W}_{ij}$ were calculated and the
Chi-squared test was performed on the reduced dataset. This sequential process is
terminated when the distribution of the reduced 
dataset is found to fit a Poisson distribution, or once as many as $20$ data points 
($10\%$) have been removed. Through this process, we determine the number of 
links that are Poisson distributed over at least the bulk ($\geqslant90\%$)
of the population. These links, together with the links 
fitting Poisson distribution over entire population, are considered to
comprise the ``representative'' structural connectivity for a human brain, 
with connections weights being $\lambda_{ij}$.

\subsection{Generating surrogate ensemble of finite size populations of brain 
networks}

In order to quantify the role of finite size effects and the specific 
distribution of $\lambda_{ij}$ values in the observed deviation of some of
the links from the Poisson distribution, we created a surrogate ensemble of
$1000$ populations, each population containing the same number ($196$) of
structural connectomes as in the empirical dataset. For each link $(i,j)$ in a surrogate connectome,
the link weight was drawn from the Poisson distribution 
$\mathcal{P}(\lambda_{ij})$ for which we used the \verb|poissrnd| function in {\em
MATLAB Release 2010b}. For each population, 
we then determined the fraction of links that deviated from
the Poisson distribution using the Chi-squared test described above. The
distribution of the fraction of deviating links $f_{dev}$ 
provides a measure of the extent to which apparent deviation of 
the link weights from
Poisson distribution, may arise from finite size effect.

\subsection{Simulated functional connectivity obtained via dynamical model for
neural population activity}
In order to obtain the functional connectome resulting from the dynamics of the structural brain network,
we use the Wilson-Cowan (WC) neural mass 
model~\cite{Wilson1972, Destexhe2009} to
describe the activity in each node of the structural connectome. 
The temporal evolution of the mean activity of excitatory ($u_i$) and
inhibitory ($v_i$) subpopulations of node $i$ is given as:
\begin{equation}
\begin{split}
\tau_{u} \dot{u}_{i} &= -u_{i} + (\kappa_u - r_{u}u_i)\mathcal{S}_{u}(u^{in}_i)\,, \\
\tau_{v} \dot{v}_{i} &= -v_{i} + (\kappa_v - r_{v}v_i)\mathcal{S}_{v}(v^{in}_i)\,, \\
\end{split}
\label{WC}
\end{equation}
where
$u^{in}_i=c_{uu}u_i - c_{uv}v_i+\sum'(w^{uu}_{ij}-w^{uv}_{ij}) + I^{ext}_u$
and 
$v^{in}_i=c_{vu}u_i - c_{vv}v_i+\sum'(w^{vu}_{ij}-w^{vv}_{ij}) + I^{ext}_v$
represent the total input to the excitatory and inhibitory subpopulations
respectively. Here, $c_{\mu\nu}(\mu,\nu=u,v)$ represents the interaction
strengths within and between the subpopulations of a node while
$\tau_{u,v}$ and $I^{ext}_{u,v}$ correspond to the time constants and the
external stimuli for each of the neural subpopulations. The interaction
strength between the subpopulations of different nodes are represented by
$w^{\mu\nu}_{ij}(\mu,\nu=u,v)$, which are obtained from the connection
weights of the structural connectome of each individual. We assume that all
inter-nodal interactions are of equal strength:
$w^{\mu\nu}_{ij}(\mu,\nu=u,v)=w_{ij}$. The summation $\sum'$ is performed
over all neighbors of the structural network. The sigmoidal response function
$\mathcal{S}_{\mu}(z)=[1+\exp\{ -a_{\mu}(z-\theta_{\mu}) \}]^{-1} + \kappa_{\mu}-1$ has a maximum value 
$\kappa_{\mu}=1-[1+\exp(a_{\mu}\theta_{\mu})]^{-1}$. Parameters are chosen
such that the dynamics of isolated nodes ($w^{\mu\nu}_{ij}=0$) are in the oscillatory
regime~\cite{Singh2016,Sreenivasan2017}. The matrices corresponding to the inter-nodal coupling weights
$w_{ij}$ are taken to be scalar multiples of individual structural weight
matrices $W$ for one set of simulations and individual rescaled weight
matrices $\mathcal{W}$ for another set of simulations. We also used other
structural connectivity matrices such as the adjacency matrix corresponding to
the rescaled weight matrices $\mathcal{A}$ and 
the two types of 
representative structural network
matrices $\langle W \rangle$ and $\Lambda$. The corresponding functional
connectivity was obtained by finding Pearson's correlation coefficient between the time
series of $u_i$ and $u_j$ for each pair of nodes $(i,j)$. For the sake of
comparison, we multiplied all the structural connectivity matrices
($W$, $\mathcal{W}$, $\mathcal{A}$, $\langle W \rangle$ and $\Lambda$) 
with corresponding normalizing constants, such that the mean connection
strength averaged over each network was always a constant $w^{avg}$. We
fixed the value $w^{avg}=100$ since at this value of average coupling we
obtained temporal activity that is qualitatively very similar to typical
empirically observed fMRI time series.

\subsection{Bimodality coefficient}

The bimodal nature of a probability distribution can be characterized by 
calculating its bimodality coefficient~\cite{pfister2013}:
\begin{equation}
 BC = \frac{m^2_3+1}{m_4+3\cdot\frac{(n-1)^2}{(n-2)(n-3)}},
\end{equation}
where $m_3$ is the skewness, $m_4$ is the excess kurtosis and $n$
represents the sample size. A distribution is considered to be bimodal if
$BC > BC^{*}$, where $BC^{*} = 5/9$. This benchmark value corresponds to a
uniform distribution, and if $BC < BC^{*}$ the distribution is considered
unimodal.	

\subsection{Statistics}
\noindent
The {\em Kernel smoothened density function}~\cite{bowman97} has been used
to estimate the probability distribution functions of different quantities
(e.g., the joint probability between $f_{ij}$ and $W_{ij}$). For this
purpose we have used the \verb|ksdensity| function in {\em MATLAB Release
2010b} with a Gaussian kernel.\\

\noindent
The {\em Two-sample Kolmogorov-Smirnov (KS) test}~\cite{Massey1951} has
been used to compare between pairs of samples (e.g., matrix correlations of
the functional and structural connectomes) in order to determine whether both of them
are drawn from the same continuous distribution (null hypothesis),
or if they belong to different distributions. For this purpose we have 
used the \verb|kstest2| function in {\em MATLAB Release 2010b}, with the
value of the parameter $\alpha$ which determines threshold significance
level set to $0.01$.

\section{Results}

\subsection*{``Wiring'' and ``weighting'' are not independent processes.}
\begin{figure*}[tbp]
\centering
\centerline{\includegraphics[width=0.85\linewidth]{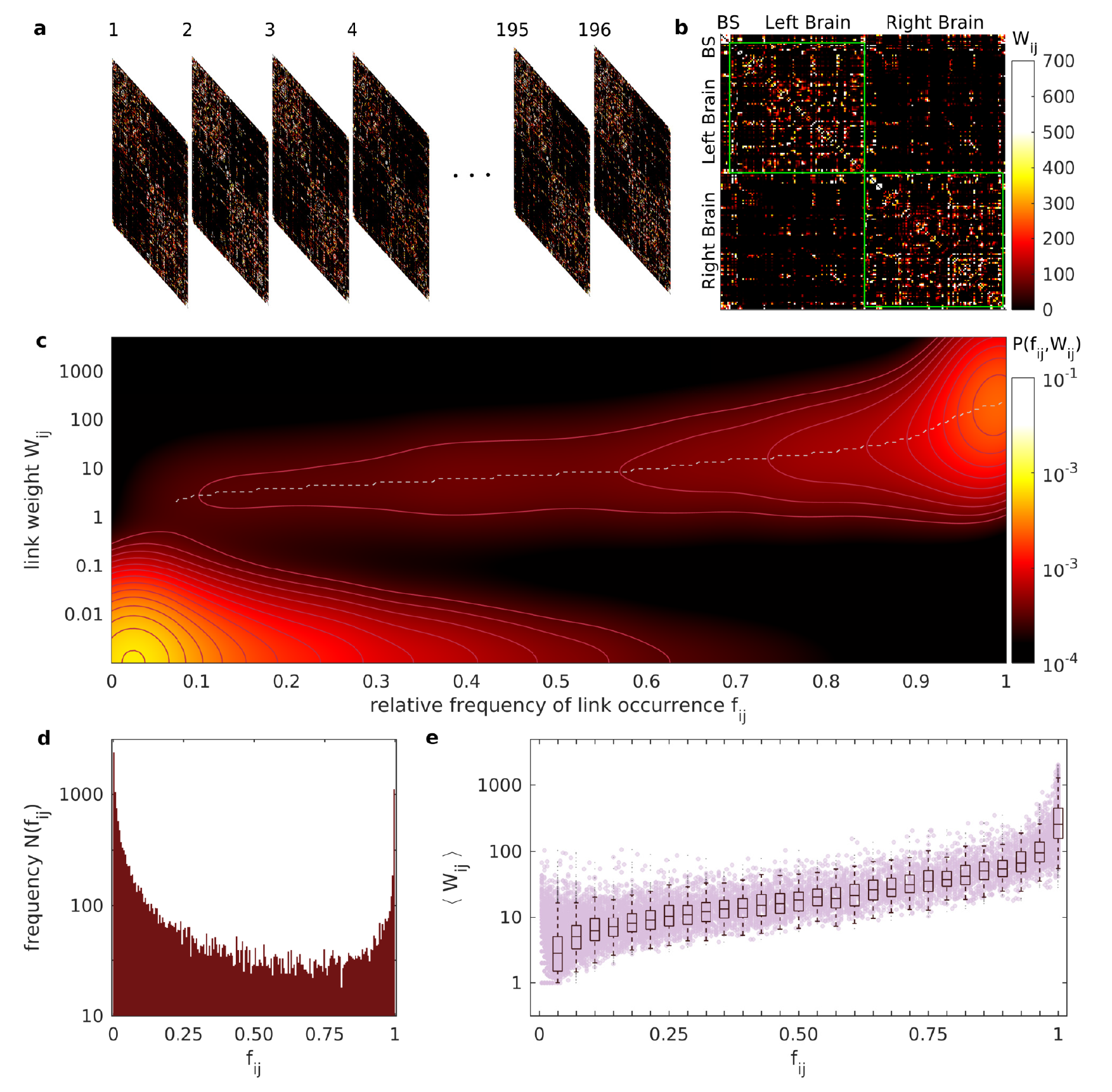}}
\caption[An increase in the occurrence of non-ubiquitous connections within
a population leads to a steady shift in the corresponding link weight
distribution towards higher values.]{
\textbf{An increase in the occurrence of non-ubiquitous connections within
a population leads to a steady shift in the corresponding link weight
distribution towards higher values.}
(a) Ensemble of weighted structural connectivity (SC) matrices representing
the structural brain networks of 196 human subjects obtained via diffusion
tensor imaging (DTI). (b) A sample SC matrix corresponding to the
connectivity information obtained from one of the subjects.
Each network comprises 188 nodes that correspond to brain regions connected
through white matter tracts, which are represented by weighted undirected
links in the SC matrix. For each network the matrix entries $W_{ij}$
represent the density of axonal bundles between nodes $i$ and $j$. The
minimum possible link weight is $1$, and matrix entries are set to $0$ if
the connection does not exist or cannot be detected due to extremely low
thickness.The regions are alphabetically arranged and grouped into the
brain-stem (BS), the left brain and the right brain. Notice the relatively
high density of ipsilateral connections (connections between regions of the
same hemisphere, as indicated by diagonal blocks in the matrix) and low
density of contralateral connections (connections between regions of
opposite hemispheres, as indicated by off-diagonal blocks). (c) Joint
probability distribution $P(f_{ij}, W_{ij})$ of the relative frequency
$f_{ij}$ that a link between brain regions $i$ and $j$ is seen across
individuals in the population and the weight $W_{ij}$ of the link. 
If a link does not exist we set $W_{ij}=10^{-3}$ and subsequently use
kernel smoothing to obtain the distribution $P(f_{ij}, W_{ij})$. We observe
a steady increase in the mode of the distribution of link weights
(represented by the broken white curve) as the frequency of occurrence
$f_{ij}$ increases, with a much steeper rise after $f_{ij}=0.9$.
Furthermore, the distribution broadens on increasing $f_{ij}$.
(d) Frequency histogram showing the distribution of links over $f_{ij}$
from the set of all possible $^{188}C_2 = 17578$ pairs of nodes $(i,j)$,
illustrating the variability in connection topology of brain networks
across the population.
(e) The mean link weights ($\langle W_{ij} \rangle$) for non-ubiquitous
links of any given frequency of occurrence $f_{ij}$, averaged over the
sub-population in which they occur, is observed to vary over an order
of magnitude with the interval of the range shifting upwards with the
increase in occurrence. Each point of the scatter plot represents a link,
which provides a link-wise resolution to the distribution of panel (c).
The box plots representing the distributions of $\langle W_{ij} \rangle$
over consecutive intervals of $f_{ij}$ clearly illustrate a steady increase
of the distribution.
}
\label{fig1}
\end{figure*}

The structural connectivity of a human brain displays large variability
across individuals in a population, in terms of both connection topology
and weight distributions. However, not all the links in the structural
network exhibit the same extent of variability. In our analysis, we
consider an ensemble of $196$ structural connectivity
(SC) matrices, as illustrated in in Fig.~\ref{fig1}~(a), which allows us to
study this variability across a diverse human population (see Methods for
details about the connectivity data). Each network contains $188$ nodes,
and the density of axonal bundles between regions $i$ and $j$ is
represented by the connection strength $W_{ij}$. Fig.~\ref{fig1} (b) shows
the SC matrix for one of the individuals, where the regions are ordered
alphabetically and grouped into three broad regions: brain-stem (BS), left
brain and right brain. Not surprisingly, we observe a considerably higher
density of ipsilateral connections than contralateral ones. In
order to characterize the topological variability of the network across the
population, for every link $(i,j)$ we measure the relative frequency of
occurrence $f_{ij}$, which is given by the fraction of the total population
in which an axonal tract between regions $i$ and $j$ is observed. For all
the $^{188}C_2$ node pairs that can in principle have a connection we
obtain the $f_{ij}$ values between $0$ to $1$, where $0$ signifies those
links that are never observed in any individual and $1$ identifies
links that are found in every member of the population. The connection
topology is largely determined during the course of development through the
process of wiring, where a complex cascade of genetic and molecular
mechanisms determine the probability of connection between two neurons. 
However, there remains uncertainty as to the exact
processes that determine the connection weights at the level of the brain
regions that are connected through axonal tracts. Note that the connection
weights in this case are not equivalent to the weights of synaptic
connections, whose strengths are governed by various learning and
plasticity mechanisms. Here the connection weights are actually the
observed density of axonal bundles, which is partially related to
physical thickness of the connections. 

\noindent
Hence there is no prior reason to
expect any correspondence between the ubiquity of a link in
the population, as quantified by $f_{ij}$, and the distribution of
its weight across the population.

Fig.~\ref{fig1}~(c) shows that as the occurrence of links in the population
increases, their weights tend to steadily rise, as indicated by the joint
probability distribution $P(f_{ij},W_{ij})$ obtained using kernel
smoothing (see Methods). The lower plateau represents the links 
with weight $W_{ij}=0$ and the upper plateau, which includes link weights from $1$ up
to order of $1000$, not only displays an increase of $P(f_{ij},W_{ij})$
with increasing occurrence (which trivially leads to increase in non-zero
weights), but also shows a steady increase of $W_{ij}$. This indicates that
those connections which are found more often in a population are likely to
have higher connection weights as well. Fig.~\ref{fig1}~(d) displays the
frequency histogram of $f_{ij}$ values, showing the highest concentration
of links at $0$ and $1$. For a species with rigidly invariant topology over
a population, such as {\em C. elegans}, this histogram would show occupancy
only at $0$ and $1$. Thus a large number of links which occur with frequencies 
$0$ and $1$ implies high topological
variability in the structural connectivity within a population. Even though
the distribution of connection weights in the population shows
correspondence with the relative frequency of occurrence $f_{ij}$,
the distribution widens with an increase in the occurrence. This implies a
large variability in connection weights, even for links with similar values
of occurrence. Assuming that the processes determining weights are still
largely independent of the processes determining the formation of the
connection itself (wiring), it is meaningful to consider the weight
distribution of non-ubiquitous links ($f_{ij}<1$) over only those
individuals from the population in which the link occurs. 
Fig.~\ref{fig1} (e) shows that even when we consider mean weights of
each link ($\langle W_{ij} \rangle$) averaged over the 
corresponding subset of
the population in which the link occurs, there is a steady rise in
these partially averaged values of the link weights. This further indicates
that wiring and weighting processes for the links are not entirely independent, 
even if they are separate processes.

\subsection*{The variation of the weights of frequently occurring links
over the population, as well as their frequency of occurrence, can both be
described by a single link-specific Poisson process.}

One of the simplest stochastic processes that describes the
distribution of the number of recorded events is the Poisson process.
It corresponds to a probability distribution of discrete random independent
events occurring at a constant rate over time or space. Here we consider the hypothesis that
the link weights $W_{ij}$ for link $(i,j)$ are generated by such 
discrete independent events occurring at a constant rate $\lambda_{ij}$
for each human subject. In such a case, the link weights
$W_{ij}$ would follow a Poisson distribution. As we have
already shown above that the processes determining the wiring and the
weights of the links appear to be at least mutually dependent, even if they
are separate, weights having Poisson distribution would actually mean that
a single parameter $\lambda_{ij}$ would be sufficient to explain the
variability of link weights as well as their frequency of occurrence.

\begin{figure*}[tbp]
\centering
\centerline{\includegraphics[width=0.85\linewidth]{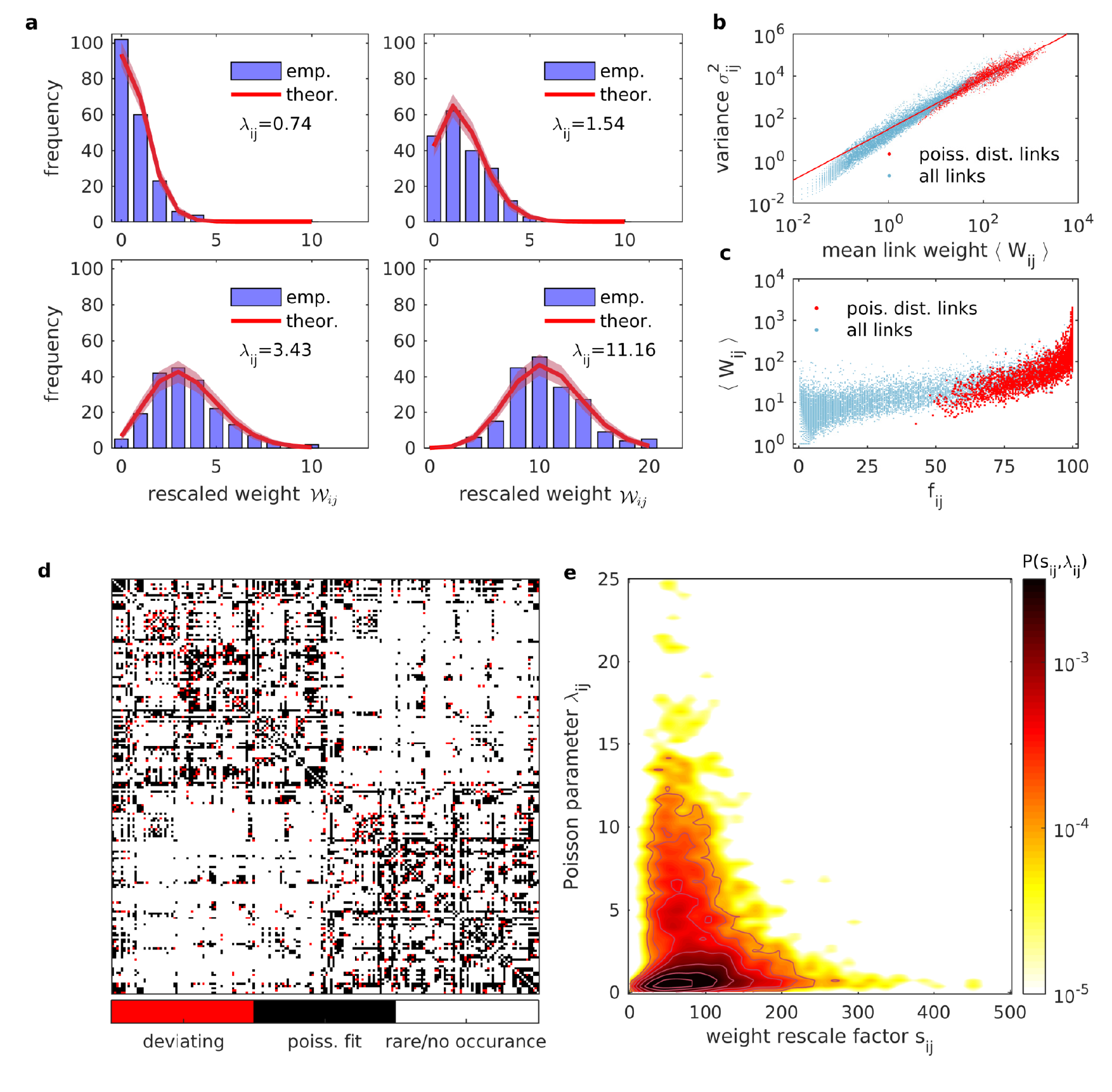}}
\caption[The variation of the weights of frequently occurring links over
the population, as well as their frequency of occurrence, can both be
described by a single link-specific Poisson process.]{
\textbf{The variation of the weights of frequently occurring links over the
population, as well as their frequency of occurrence, can both be described
by a single link-specific Poisson process.}
(a) Frequency histograms corresponding to four separate links $(i,j)$
demonstrating that their rescaled link weights $\mathcal{W}_{ij}$, obtained
using corresponding rescaling factors $s_{ij}$, are distributed over the
population in a Poisson process, with respective Poisson parameters
$\lambda_{ij}$ displayed in each case.
The solid lines, which represent the theoretical values of the
distributions, display close agreement with the histograms corresponding to
the observed distributions and the shaded region represents
the fluctuations in the frequencies over a large number of randomly drawn
samples of size $196$ (same as the population size) from the corresponding
Poisson distribution. The goodness of fit for the links with the Poisson
distribution has been quantitatively tested using Pearson's Chi-square
test~\cite{Pearson1900}.  
(b) Scatter plot of the links, represented in terms of their mean weights
across entire population $\langle W_{ij} \rangle$ and the their variances
$\sigma^2_{ij}$. It can be seen that for most links the weight
distributions have $\sigma^2_{ij} \propto \langle W_{ij} \rangle$, which
indicates the possibility of a rescaled Poisson distribution. The links that
fit a rescaled Poisson distribution (as shown by Chi-squared
goodness of fit test with significance $\alpha=0.01$), are distinguished by
showing them in red color and the linear regression fit for those points
(red line) has a slope of $1.2$. (c) Links which are shown to fit rescaled
Poisson (red dots) have high values for the relative frequency of
occurrence ($> 0.5$) as well as higher values of mean link weights.
(d) Adjacency matrix showing the entries corresponding to $3342$ links that
fit the rescaled Poisson distribution with $\alpha=0.01$ (black entries)
and $577$ links whose weight distributions deviate from the corresponding
Poisson distributions (red entries). (e) Joint probability distribution
$P(s_{ij}, \lambda_{ij})$ of the distribution of rescaling factors $s_{ij}$
and Poisson parameter $\lambda_{ij}$. We observe that $P(s_{ij},
\lambda_{ij})$ is relatively unaffected by increasing or decreasing
$\lambda$, which suggests that both these link properties that together
determine the observed weights for a link in the population, might
originate from distinct biological factors.
}
\label{fig2}
\end{figure*}

\noindent
We find that for a large number of frequently occurring links
($f_{ij}>0.5$), their weights follow a rescaled Poisson distribution
(see Methods) with each link having specific value of $\lambda_{ij}$ and
corresponding rescaling factor $s_{ij}$. The rescaling factor is a
link-specific constant scalar value for a link $(i,j)$, which can be applied on 
its weights $W_{ij}$ across the population to obtain
the rescaled weights $\mathcal{W}_{ij}$, which follow a Poisson
distribution $\mathcal{P}(\lambda_{ij})$ (see Methods for
details). Using Pearson's Chi-squared goodness of fit test (see Methods) to
determine whether the rescaled weights for a link fits a Poisson
distribution with high significance ($\alpha=0.01$), we find that
out of $15,209$ links that are seen at least once in the population there
are $3342$ links whose rescaled link weights can be described by Poisson
processes and $11,290$ links have too few occurrences for a reliable
statistical fitting. Thus there are only $557$ links that deviate significantly from the
Poisson distribution.

Fig.~\ref{fig2}~(a) demonstrates that the frequency histogram
corresponding to rescaled link weights of four separate links agree
with the theoretically expected frequencies from a Poisson distribution having the 
corresponding $\lambda_{ij}$. As both the mean and variance of Poisson
distributions are equal to the same parameter $\lambda$, it is expected that the rescaled
Poisson distribution followed by $W_{ij}$ has
$\langle W_{ij} \rangle \propto \sigma^2_{ij}$, as can be seen in
Fig.~\ref{fig2}~(b). Fig.~\ref{fig2}~(c) shows that links described by a
Poisson process appear to have high significance in the structural
connectivity in terms of both the topology (by having higher values of 
$f_{ij}$), as well as, the strength of connection (by having higher
values of partially averaged mean link weights as described in
Fig.~\ref{fig1}~(e)). The adjacency matrix in Fig.~\ref{fig2}~(d)
illustrates the highly dense network comprising of links described by
Poisson processes, while also showing the sparsely distributed $557$
links that deviate from the Poisson process. Further, the two factors
determining the connection weight for a link, viz. $\lambda_{ij}$ and 
$s_{ij}$, seem to have no dependence on each other, as can be seen in their
joint probability distribution in Fig.~\ref{fig2} (e). The Pearson's
correlation coefficient between respective values of $\lambda_{ij}$ and $s_{ij}$ 
is $-0.09$.

\begin{figure*}[tbp]
\centering
\centerline{\includegraphics[width=0.85\linewidth]{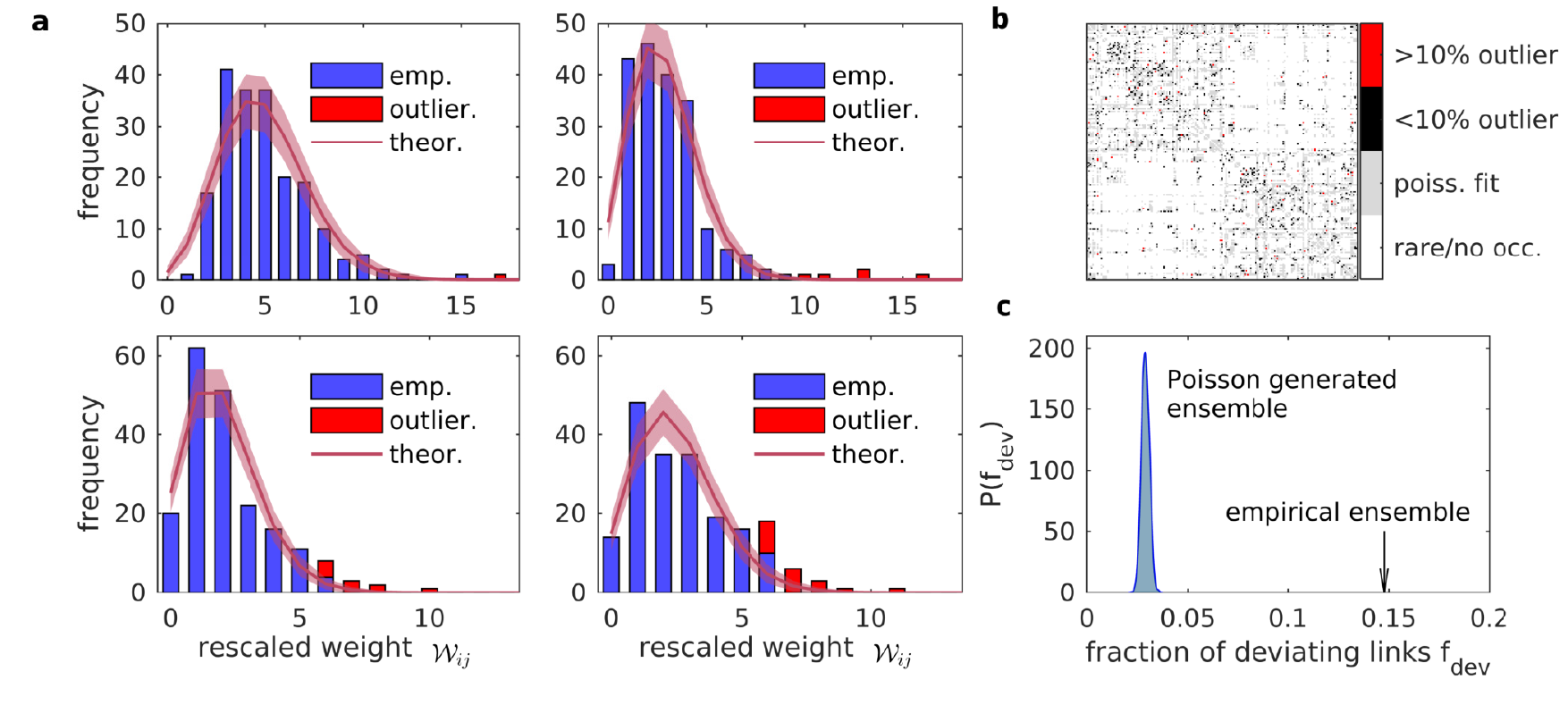}}
\caption[Exclusion of a small fraction of outliers can make deviating links fit a Poisson distribution.]{
\textbf{Exclusion of a small fraction of outliers can make deviating links fit a Poisson distribution.}
(a) Frequency histograms corresponding to four separate links $(i,j)$
demonstrating that rescaled link weights $\mathcal{W}_{ij}$ obtained using
corresponding rescaling factors $s_{ij}$ are distributed over the bulk of the
population in a Poisson process (represented by blue bars),
upon excluding a few outliers ($<10\%$ of the population, represented by
red bars) with high values that deviate from the Poisson process. The solid
lines which represent the theoretical values of the distributions display
close agreement with the histograms corresponding to the observed
distributions for the bulk of the population (blue bars) and the
shaded region represents the variability in the frequencies
across a large number of randomly drawn samples of size $196$ (same as the
population size) from the corresponding Poisson distribution.
(b) Adjacency matrix showing that $520$ links (shown as black entries) out
of the $577$ links that originally deviated from the Poisson distribution
over the entire population (as shown in Fig.~\ref{fig2} (d)), fit the
Poisson distribution on removing only upto $10\%$ outliers from the
population, as quantitatively shown using the Chi-squared test.
The remaining $57$ links do not fit a Poisson distribution even after the
removal of $10\%$ outliers (shown as red entries). (c) The empirically
observed fraction of links whose weights deviate from the Poisson
distribution $f_{dev}$ (out of all the links with sufficient frequency of
occurrence, viz.$\approx 0.15$, as indicated by the arrow), cannot be
explained by finite size effect alone. The probability
distribution shows the expected values for $f_{dev}$ in a randomly
generated network ensemble of sample size $196$, where each link weight
across all the networks is drawn from a Poisson distribution with
link-specific values of $\lambda_{ij}$ that were obtained from the
empirical dataset. Finite size effects contribute a value of $f_{dev}\approx 0.03$
which is much smaller than that observed
from empirical data.
}
\label{fig3}
\end{figure*}

We next examine the $577$ links which, in spite of having high frequency
of occurrence, do not fit Poisson distribution. We observed that the bulk
of these links have a bimodal distribution of rescaled weights, in which the
first mode appears to be a Poisson distribution while the other
mode (occurring at much higher values) comprises outliers with unexpectedly high
weights. Hence the deviation from a purely Poisson process can possibly be
attributed to such outliers. The rescaling factor $s_{ij}$ and
Poisson parameter $\lambda_{ij}$ obtained from these links are possibly 
inaccurate because they are calculated by including the outlier values. 
Thus, for each of
these deviating links, we sequentially remove the data points with 
the largest weight, calculate the $\lambda_{ij}$ and $s_{ij}$ for the remaining
subset and rescaling the weights until the remaining data agrees with the
Poisson process as per Chi-squared goodness of fit with a high significance
($\alpha=0.01$). Out of $577$ deviating links, we find that $520$ links
fit the rescaled Poisson distribution after excluding less than $10\%$ of
the outliers. Fig.~\ref{fig3} (a)
demonstrates that in four of the deviating links that are shown, the
bulk of the distribution ($>90\%$) agrees with a Poisson distribution,
while the outliers ($<10\%$) deviate significantly from the theoretical
distribution. The adjacency matrix in Fig.~\ref{fig3}~(b) shows that
the bulk of the deviating links fit the Poisson distribution by excluding few outliers.
We note that out of a total of $3919$ links that have sufficient
occurrences for statistical tests, $3862$ links ($98.5\%$) have their
weights distributed according to link specific Poisson processes for
more than $90\%$ of the population.

The deviation observed in these $577$ links suggests that even though a
single Poisson process might be sufficient to describe the wiring and the
weights for a large fraction of links for most of the population, there
may be other factors in the generative mechanism that lead to a
significant fraction of links ($\approx0.15$) deviating from a Poisson
process, even if that deviation is due to the presence of a few outliers within the
population. In order to rule out the possibility that the deviation arises simply because of 
finite size effects in the specific
distribution of $\lambda$ values, we generated a surrogate ensemble of SC
matrix sets, each comprising $196$ matrices. The link weights were each
drawn from the link-specific Poisson distributions (see Methods for
details). We find that in every realization, a small fraction of links
does not fit a Poisson distribution, as per the Chi-squared criterion.
The distribution of these values is shown in Fig.~\ref{fig3}~(c).
The deviations arising due to such finite size effects are significantly lower
than those seen empirically. This suggests that there must be other
significant factors in the generative mechanisms apart from the Poisson
process that give rise to the structural connectivity in human brains.

\subsection*{Rescaling of link weights using Poisson parameters might
provide greater functional interpretability to the structural
connectivity.}

While we have have thus far interpreted the Poisson parameter
$\lambda_{ij}$ for the links as an abstract representation of the net effect
of all biological factors associated with the generative mechanism of the
structural connectivity of brain networks, an additional interpretation of
the Poisson parameter can be obtained by considering its effect on
function. While there have been several attempts to examine the
relation between structural connectivity and functional
connectivity~\cite{Honey2009, Hagmann2008}, it has been observed that
there is very low correspondence between the two. Unlike synaptic
and gap-junction weights which are interpreted as coupling strengths
between neuronal activities, the connection weights in the structural macro-connectome
elude a functional interpretation (such as for instance
being a measure of dynamical coupling between the activities of
different brain regions). We have already observed that rescaled weights
appear to be a more fundamental structural property than the original
weights, since they are essentially discrete Poisson variables which are
known to arise in a wide range of natural 
phenomena. It is therefore reasonable to ask whether the rescaled
weights can have a deeper relation with the functional coupling.
For example, could the rescaled weight be directly related to the number of
axons in axonal tracts? Here we ask whether the individual structural
connectivity of rescaled weights, which are obtained from the
analysis of SC matrices across a population, have a stronger correlation
with the corresponding functional connectivity as compared to
the original structural connectivity.

\begin{figure*}[tbp]
\centering
\centerline{\includegraphics[width=0.85\linewidth]{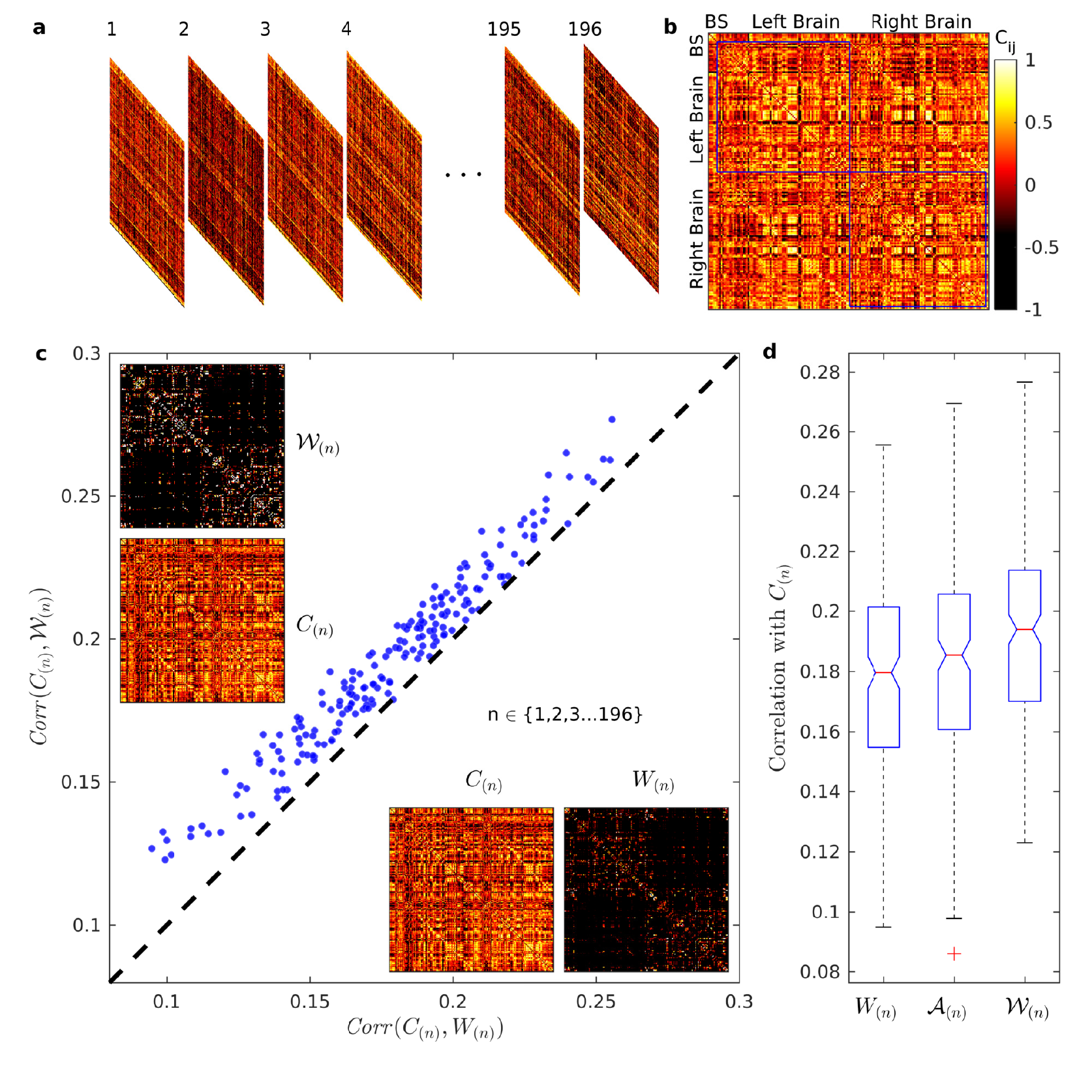}}
\caption[Upon rescaling the weights, the structural connectivities
consistently show a greater association with the corresponding functional
connectivities across the population.]{
\textbf{Upon rescaling the weights, the structural connectivities
consistently show a greater association with the corresponding functional
connectivities across the population.}
(a) Ensemble of functional connectivity (FC) matrices representing the
resting state functional brain networks, determined by functional magnetic
resonance imaging (fMRI), of the same $196$ human subjects whose structural
connectivities have been analyzed in this work.
(b) A sample FC matrix corresponding to one of the subjects. The nodes are
arranged in the same way as described in Fig.~\ref{fig1} (b), and the
matrix entries $C_{ij}$ indicate the correlation between the haemodynamic
activities of nodes $i$ and $j$ over a period of time. Notice the striking
difference between the SC shown in Fig.~\ref{fig1} (b) and the FC shown
here. Where the SC have relatively sparse connections and a much higher
density of ipsilateral connections than the contralateral ones, the FC have
a much higher density of strong connections (both positive and negative)
with equal density of functional connectivity between ipsilateral (diagonal
blocks) as well as between contralateral regions (off diagonal blocks).
(c) A comparison of  the dependence between individual functional
connectivity matrices ($C_{(n)}$) for an individual $n$ (where
$n \in {1, 2, 3 \ldots 196}$) and the corresponding structural
connectivities described respectively by weight matrices ($W_{(n)}$) and
rescaled weight matrices ($\mathcal{W}_{(n)}$). Examples of the three types
of matrices are shown as insets. The scatter plot displays the correlation
between $C_{(n)}$ and $W_{(n)}$ for each $n$ along the x-axis, and the
corresponding correlation between $C_{(n)}$ and $\mathcal{W}_{(n)}$ along
the y-axis.  Note that while the correlations between functional
and structural matrices of both types are very low ($<0.3$) (even though
statistically significant), for every individual in the population the
functional connectivity matrices are more correlated with the corresponding
$\mathcal{W}_{(n)}$ than they are with $W_{(n)}$. (d) Box-plots
representing the distributions of the correlations of the functional matrix
$C_{(n)}$ for each individual $n$ with the corresponding matrices weight
$W_{(n)}$ and $\mathcal{W}_{(n)}$ and the adjacency matrix
$\mathcal{A}_{(n)}$, generated from the rescaled weight matrix
$\mathcal{W}_{(n)}$ by removing information about the link weights. The
correlation distribution for $\mathcal{W}_{(n)}$ which is positioned higher
than that corresponding to $W_{(n)}$ as expected from the result in panel
(c), is higher compared to the correlation distribution for
$\mathcal{A}_{(n)}$, which contains information related to the topology of
the rescaled matrix but not the link weights. Note that topology of the
rescaled weight matrices are different from that of the original weight
matrices, since the former has much fewer links. This implies that in
addition to the altered topology of rescaled weight matrices, the altered
weight distribution also makes $\mathcal{W}_{(n)}$ a better structural
correlate of the observed dynamical function of the brain.
}
\label{fig4}
\end{figure*}

The {\em Nathan Kline Institute (NKI) / Rockland Sample} dataset that we 
consider~\cite{Nooner2012} includes resting state functional connectivities
(FC) for each of the $196$ individuals whose structural connectivity we
have analyzed thus far. Sample functional connectivity matrices are
displayed in Fig.~\ref{fig4}~(a). In contrast to the structural
connectivities, the functional connectivities exhibit a large degree of variability
across individuals. Examining a single FC matrix (Fig.~\ref{fig4}~(b))
reveals that the density of strong functional connections is much higher
than that of structural connections (Fig.~\ref{fig1}~(b)), and unlike the
structural networks there is a significant number of connections across
left and right hemispheres. We first examine the correlation between the
structural and functional connection strengths for each link ($W_{ij}$ and
$C_{ij}$, respectively) over the entire population and observe that most of
the links have negligible correlation between the variations of structural
and functional connectivity strengths over the population (see Fig. S1 in SI).
While this supports the previously known observation that there is very low correspondence
between structure and function at the level of individual links, we observe
that a macroscopic comparison between the FC matrices ($C_{(n)}$) and structural weight
matrices ($W_{(n)}$) for each individual $n$ reveals statistically
significant correlations, as seen in Fig.~\ref{fig4}~(c). Furthermore, for
every individual in the population we observe that the matrix correlation
of $C_{(n)}$ and structural weight matrix $W_{(n)}$ is lower than the matrix correlation of
$C_{(n)}$ and the corresponding rescaled structural weight matrix $\mathcal{W}_{(n)}$.
Thus, there is a notable correspondence between

\noindent
structure and function for
each individual at the level of the entire brain network, and furthermore,
the rescaling of weights in the structural connectivity enhances this
correspondence between structure and function in all individuals. A
possible explanation for this enhanced correspondence with function might be the alteration in the distribution of
weights after rescaling, or it might also be attributed to the change in the connection topology due to the rescaling of
weights (as the rescaling process leads to the deletion of links). To
examine this, we computed matrix correlations of each $C_{(n)}$ with
(unweighted) adjacency matrices having the same connectivity as
$\mathcal{W}_{(n)}$, which we refer to as $\mathcal{A}_{(n)}$. The boxplots
in Fig.~\ref{fig4}~(d) shows that the matrix correlations of $C_{(n)}$ with
$\mathcal{W}_{(n)}$ are higher than correlations with $\mathcal{A}_{(n)}$,
suggesting that the alteration of the weight distribution upon rescaling is
the primary factor in making SC a better correlate of FC. The three
distributions of matrix correlation values were found to be significantly
distinct from one another using the {\em Two-sample Kolmogorov-Smirnov (KS)
test} (see Methods).

\begin{figure*}[tbp]
\centering
\centerline{\includegraphics[width=0.85\linewidth]{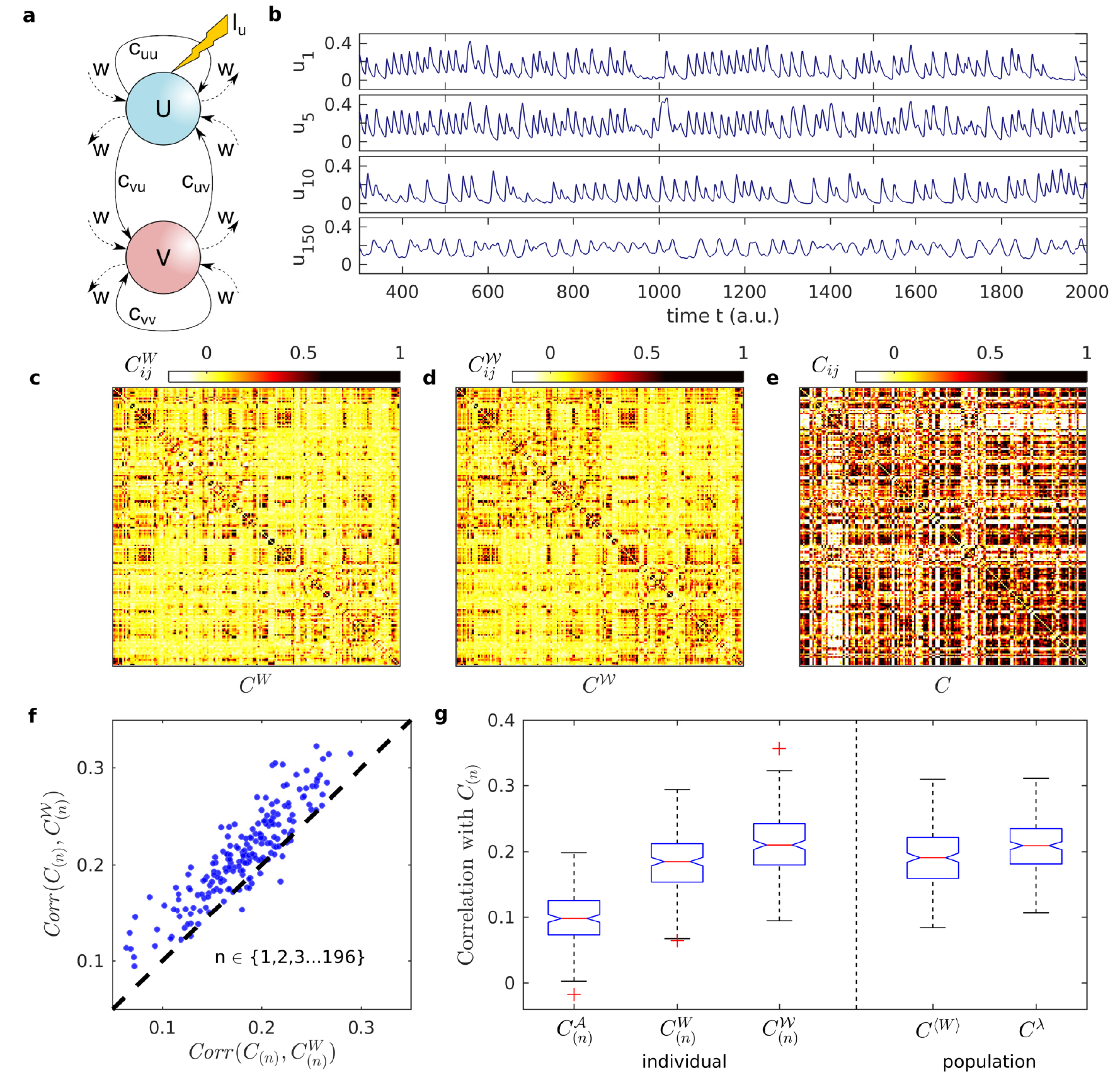}}
\caption[Dynamical simulation of whole brain activity using a neuronal
population model suggests that structural connectivity with rescaled
weights are better structural correlates for brain function than those with
original weights.]{
\textbf{Dynamical simulation of whole brain activity using a neuronal
population model suggests that structural connectivity with rescaled
weights are better structural correlates for brain function than those with
original weights.}
(a) Schematic representation of a single dynamical element (oscillator) of
the Wilson-Cowan (WC) model, which simulates the activity within a single
region of the brain network, each comprising
interactions between the excitatory and inhibitory subpopulations ($U$ and
$V$ respectively), with strengths denoted by $c_{\mu\nu}$ where
$\mu,\nu=\{u, v\}$, and their interactions with the subpopulations
belonging to other nodes of the network (with uniform coupling strength
$w$). The lightning bolt represents the external stimulation of strength
$I_{u}$ provided to the excitatory subpopulation. (b) Time evolution of the
dynamical activity of the excitatory subpopulations in four out of $188$
brain regions of a single individual, obtained from simulations of WC
oscillators on the network specified by the structural connectivity matrix
associated with the rescaled weights $\mathcal{W}$ as coupling strengths.
As the model is dimensionless, here time is displayed in arbitrary units
(a.u.). (c-d) Functional connectivity matrices obtained from simulated
brain activity using one of the connectomes.
The matrices $C^W$ and
$C^{\mathcal W}$ are respectively obtained from simulations of WC
oscillators on (c) the original weight matrix $W$, and (d) the rescaled
weight matrix $\mathcal{W}$. (e) Empirical functional connectivity obtained
from the resting state brain activity of the same individual. (f) Scatter
plot displaying the correlation between the empirical FC matrices $C_{(n)}$
and the simulated FC matrices obtained from the original weight matrices
$C^W_{(n)}$ for each individual $n$ along the abscissa, and the correlation
between $C_{(n)}$ and the simulated FC matrix obtained from rescaled weight
matrix $C^{\mathcal{W}}_{(n)}$ along the ordinate. It can be observed that
$corr(C_{(n)}, C^{\mathcal{W}}_{(n)})$ is consistently higher than
$corr(C_{(n)}, C^{W}_{(n)})$ for the majority of individuals. This 
further extends the result shown in Fig.~\ref{fig4} (c). (g) Box-plot
showing that the correlations of the empirical FC of individual
brains $C_{(n)}$ with simulated FC from corresponding rescaled weight
matrices $C^{\mathcal{W}}_{(n)}$ are comparatively higher than both
$corr(C_{(n)}, C^W_{(n)})$ and $corr(C_{(n)}, C^{\mathcal{A}_{(n)}})$,
where $C^W_{(n)}$ represents the simulated FC matrices obtained from
corresponding weight matrices,  and $C^{\mathcal{A}}_{(n)}$ represents the
corresponding adjacency matrices of the rescaled weights. Furthermore, as
$corr(C_{(n)}, C^{\mathcal{A}_{(n)}})$ is the weakest, it suggests that the
dynamical behavior of the brain is governed more strongly by the weight
distribution of the structural connectivity than by the network topology.
We also obtain simulated FC generated from two alternative 
representative structural networks of a human brain, viz.
$C^{\langle W \rangle}$ which is generated from the matrix
$\langle W \rangle$ that we obtain by averaging each link weight $W_{ij}$
over entire population and $C^{\lambda}$ which is generated from the matrix
$\lambda$, which comprises the Poisson parameters $\lambda_{ij}$ for each
link that is Poisson distributed over at least $90\%$ of the population.
We observe that the empirical FC matrices correlate better with the
simulated FC $C^{\lambda}$ compared to the simulated FC
$C^{\langle W \rangle}$, as indicated by the box-plots.
}
\label{fig5}
\end{figure*}

The resting state functional connectivities of the individuals in the
dataset are simplified snapshots of a large repertoire of complex dynamical
behaviors in the brain, which are associated with various cognitive
functions and arise from highly complex non-linear interactions among the
constituent brain regions. Therefore a simple correspondence between
structural and functional connectivities would not be sufficient to
establish that a rescaled weight distribution in structural connectivity
would be a better functional correlate than the original weight
distribution. In order to understand how the weights in the structural
connectivity affect the dynamics, we simulate the complex dynamical
activity that would arise when the structural connectivity provides the
basis for the non-linear interactions between the brain regions. To
describe the dynamical activity of brain regions, we use a well-known
neural mass model, viz., the Wilson-Cowan (WC)
model~\cite{Wilson1972, Destexhe2009} (see Methods for details).
Fig.~\ref{fig5}~(a) shows a schematic representation of one WC node which
is used to represent the activity in a single brain region.  In order to
generate the time series of simulated activity of all brain region for each
individual, we simulated systems of WC oscillators placed at the nodes of
networks with the corresponding structural connectivity matrix ($W_{ij}$, as
well as, $\mathcal{W}_{ij}$), with the associated connection strengths of
these matrices taken to be the same as the dynamical coupling strengths
$w_{ij}$ between WC nodes. Fig.~\ref{fig5}~(b) shows samples of such time
series. By computing Pearson's correlation coefficient 
between the time series of activity for
each pair of brain regions, we generated the simulated functional
connectivity matrices $C^W$ (Fig.~\ref{fig5}~(c)) and $C^{\mathcal{W}}$
(Fig.~\ref{fig5}~(d)) for each individual to compare with the corresponding
empirical functional matrices $C$ (Fig.~\ref{fig5}~(e)). We compared the
matrix correlations between individual $C_{(n)}$ and
$C^{\mathcal{W}}_{(n)}$ with correlations between individual $C_{(n)}$ and
$C^W_{(n)}$ matrices, as shown in Fig.~\ref{fig5}~(f). Similar to the
result obtained from the comparison of FC matrices with SC matrices in
Fig.~\ref{fig4}~(c), we find that simulated FC generated from rescaled
weight matrices $\mathcal{W}_{(n)}$ were consistently better correlated
with empirical FC than the ones generated from original

\noindent
weight matrices $W_{(n)}$. We also generated simulated FC from adjacency matrices
$\mathcal{A}_{(n)}$ obtained from $\mathcal{W}_{(n)}$ and found that
FC generated from adjacency matrices had the weakest correlation
with the empirical matrices out of all three types of simulated FC, as seen
in first three box-plots in Fig.~\ref{fig5}~(g). This strongly suggests
that the dynamics in complex non-linear systems such as the brain is
governed much more strongly by the weight distributions of the connectivity
than the connection topology itself.  

Finally, we compare each of the empirical FC with the corresponding matrix
generated by a generic ``representative'' network that would describe the
structural connectivity of the human brain. There are two alternative ways to
obtain a ``representative'' network from the SC ensemble. The widely
used approach is to obtain an average network $\langle W \rangle$, which 
comprises the average weight of each link calculated over entire population. Our
results suggest a second approach, which is to consider the $\lambda$
matrix, comprising the Poisson parameters $\lambda_{ij}$ for all the
Poisson distributed links, as the representative network. By construction,
it comprises only those links that follow a Poisson distribution. As we
have already observed that rescaled weights of an individual SC might be
more relevant in interpreting structural connections, it is
more meaningful to consider the $\Lambda$ matrix (which effectively is the
average of all rescaled weight matrices), as the representative structural 
network. We generate simulated FCs corresponding to $\langle W \rangle$ and
$\Lambda$, and observe that FC generated from $\Lambda$ is more strongly
correlated with the bulk of the empirical FCs than the one generated form
$\langle W \rangle$, as seen in the last two box-plots of
Fig.~\ref{fig5}~(g). All the correlation distributions being considered
here have been shown to be significantly distinct from each other
using the {\em Two-sample Kolmogorov-Smirnov (KS) test} (see Methods).

\subsection*{The representative structural connectome can
be resolved into two components: ``Basal'' network and ``Superstructure''
network.}

\begin{figure*}[tbp]
\centering
\centerline{\includegraphics[width=0.85\linewidth]{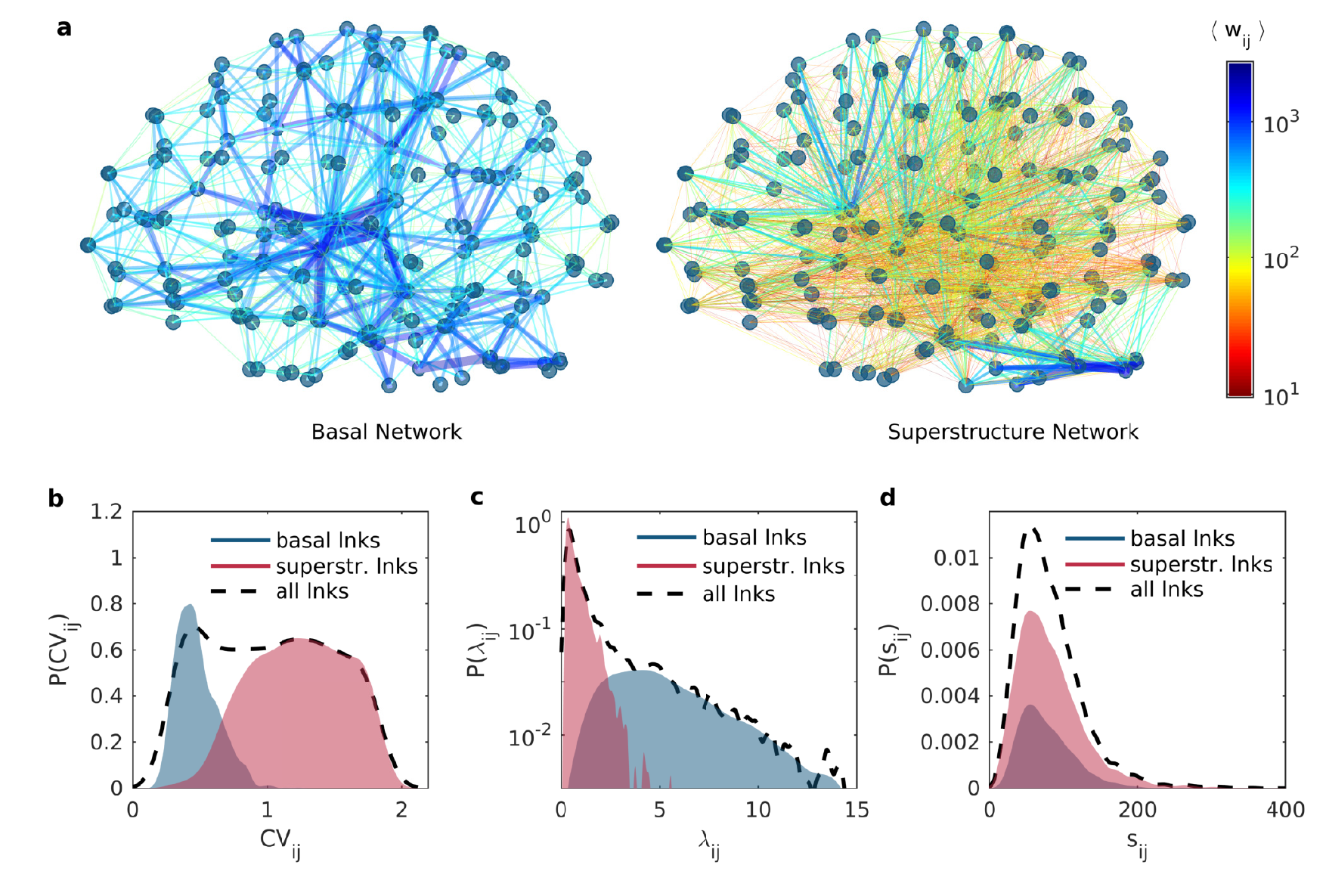}}
\caption[The representative structural connectivity of a brain network can
be resolved into two components.]
{\textbf{The representative structural connectivity of a brain network can
be resolved into two components.}
(a) Sagittal plane projections of the ``basal'' network (left)
and the ``superstructure'' network (right). The former comprises $1106$
ubiquitous links, i.e., those that occur in every individual, and the latter
consists of the remaining $2806$ links. The thickness and color of each
link between a pair of regions $(i,j)$ corresponds to their average weights
$\langle w_{ij}\rangle$ across the population (see legend). Note that the
average link weights in the basal networks are much higher than those in
the superstructure network. (b) The distribution of the coefficients of
variation $CV_{ij}$ for the link weights across the population (indicated
by broken lines) is observed to be bimodal. The mode corresponding to lower
values of $CV_{ij}$ is attributed to links from the basal network
(blue shaded region), whereas the links from
the superstructure network (red shaded
region) primarily contribute to the mode corresponding to higher values of
$CV_{ij}$. This demonstrates that the links of the basal network tend to
have higher link weights on an average, and their weights are
largely invariant across the population. In contrast, the link weights in
the superstructure network vary across individuals.
(c) The distribution of Poisson parameters $\lambda_{ij}$ (indicated by
broken lines) is also bimodal, with each mode corresponding to the basal
network links (blue shaded region) and superstructure links (red shaded region)
respectively. (d) The distribution of weight rescaling factors $s_{ij}$
(broken lines) is observed to be unimodal, in contrast to $\lambda_{ij}$
and $CV_{ij}$, with no distinction between the basal and the superstructure
links as indicated by their strongly overlapping distributions (blue and
red shaded regions). Note that the separate distributions for basal
network links and superstructure network links are normalized according to
their relative sizes.
}
\label{fig6}
\end{figure*}

\begin{figure*}[tbp]
\centering
\centerline{\includegraphics[width=0.85\linewidth]{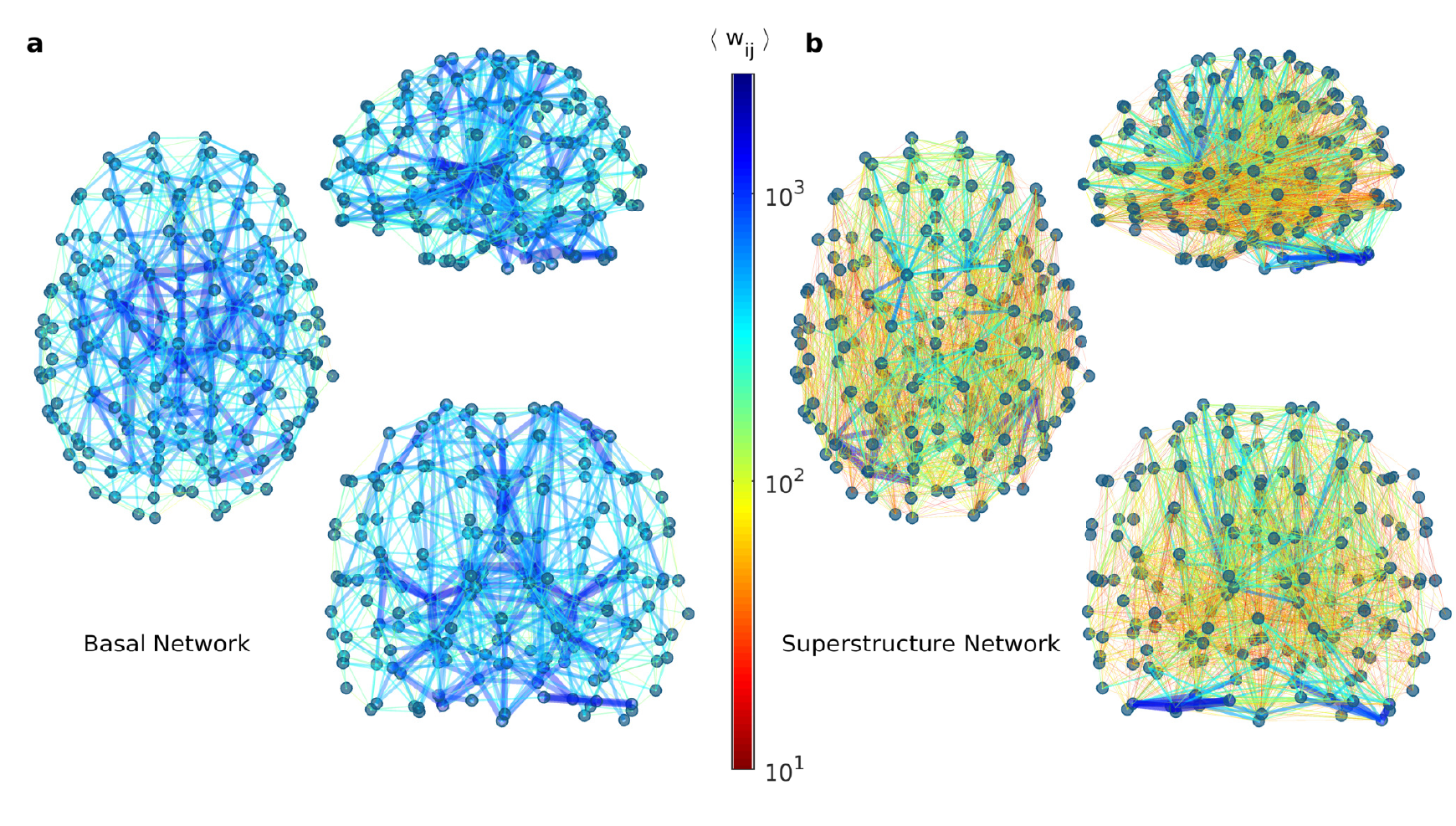}}
\caption[``Basal'' Network and ``Superstructure'' Network.]{
\textbf{``Basal'' network and ``Superstructure'' Network.}
(a) Horizontal, sagittal and coronal projections of the spatial
representations for the ``basal'' network (left, top right and bottom right
respectively). (b) Horizontal, sagittal and coronal projections of the
spatial representations for the ``superstructure'' network (left, top right
and bottom right respectively). Note that the average weights
$\langle w_{ij}\rangle$ (represented by the thickness and color of the
links $(i,j)$, see legend) display a smooth spatial gradient. In addition,
the long range connections between spatially distant regions are more
frequently observed in the superstructure network, as compared to the basal
network whose structure is closer to a lattice in that most of the
connections are between spatially adjacent regions.
}
\label{fig7}
\end{figure*}

We have thus far argued that the representative network described by the
$\Lambda$ matrix is significant because: (i) the underlying generative
mechanisms for determining the wiring and weighting of links can be
described by a Poisson process (Fig.~\ref{fig2} and Fig.~\ref{fig3}),
(ii) the constituent links are significant in terms of topology, as well as,
connection weights (Fig.~\ref{fig2}~(c)), and (iii) The $\Lambda$ matrix is
a better structural correlate to observed function in comparison
to the average SC matrix. We now return to one of our original questions
regarding the extent of variability of the structural network within the
population in terms of topology and weight distribution, focusing only on
the constituent links of the representative network. On examining how
various link-specific properties, such as the coefficient of variation of
link weights ($CV_{ij}$) and Poisson parameters ($\lambda_{ij}$) are
distributed among the constituent links of the representative network, we
find that the network can be resolved into two distinct classes of links
that we refer to as the ``basal'' and the ``superstructure'' network, shown
in Fig.~\ref{fig6}~(a). The former comprises links that are seen in all
individuals ($f_{ij}=1$) while the latter contains all the remaining links
of the representative network. They can be identified from the clearly
observable bimodality in the distributions of $CV_{ij}$
(Fig.~\ref{fig6}~(b)) and $\lambda_{ij}$ (Fig.~\ref{fig6}~(c)). The
bimodality in both distributions can be verified by calculating their
bimodality coefficients (see Methods). 
Basal links are distinguished by very low variability in weight
across the population but high values of average weights and
$\lambda_{ij}$, while the superstructure links show highly variable connection
weights across the population, but typically low values of average link
weights and $\lambda_{ij}$. Notably, the distribution of weight
rescaling factors $s_{ij}$ does not show any distinction between the basal
and superstructure networks (Fig.~\ref{fig6}~(d)).
Planar projections of the basal and superstructure networks on horizontal,
sagittal and coronal planes are provided in Fig.~\ref{fig7}. 

Using the representative network for human brain as the basis, we can explore the mesoscopic organization of human brain
in the same way as was done for macaque brain in~\cite{Pathak2020}. We have shown the preliminary results from our modular analysis
of human brain network in SI.


\section{Discussion}
	
Our results suggest that the expected weight distribution of a 
link in the structural connectome,
as well as its expected probability of
occurrence in an individual, can be described by a single parameter. This
is indicative of a common generative mechanism that determines both the
connection topology of the axonal tract wiring between brain regions, as
well as, the anatomical thickness of the tracts, which we refer to as the
connection weight. At the neuronal scale connection weights refer to the
number of synapses between two neurons, or the synaptic conductivity.
These quantities have a direct measurable effect 
on the complex electrophysiological interactions between the neurons. The 
plasticity and learning mechanisms that determine and alter the connection
weights between neurons are well understood, e.g., spike-time dependent
plasticity. However, in the case of the macro-connectome, the role of
connection weights, viz., the density of axonal bundles, in the functional
interactions of the brain areas is not well understood. By determining
a latent Poisson distributed quantity from the observed weight of a
connection, which we refer to as the rescaled weight, our results point
towards a potential framework for a better functional interpretation of
structural connectivity. This might be extremely useful in the development
of dynamical models of various cognitive phenomena in the brain. To probe
this, we have used a relatively simple neural mass
model to show that the rescaled weights 
consistently give rise to dynamical behavior that have better
correspondence with the empirical data. This mode of functional
interpretation of structural connectivity can be further enhanced when we
also consider information related to whether the synaptic connections
underlying given axonal pathways are excitatory, inhibitory or both.
Another component that is missing from the analysis of structural
connectome is the directional information about the connections. Although we represent
the structural connectome as an undirected weighted network, in reality
synaptic connections are always directed. The observation that the rescaled
weight distribution showed a widespread enhancement in correspondence
between structure and function, even without the associated information
about the directionality and the type of connections, underpins the
significance of this framework. In that case, the latent Poisson parameter
associated with the axonal pathway, i.e., the rescaled weight, might
represent functionally relevant factors such as the actual number of
axons in the tract or the number of synaptic connections, rather than
simply representing the anatomical thickness. Similarly, the corresponding
rescaling factors might be representative of peripheral features
that do not directly affect the neuronal
interactions, e.g., the thickness of myelin sheaths covering the axons.

We further observe that a significant fraction of links showed deviations
from a Poisson generative process, and these deviations are far greater than
expected by the finite size of the data. This further illuminates the generative
mechanism: even though random independent discrete processes might
be involved in the wiring of major portion of the brain, as indicated
by the occurrence of Poisson distributions, there are other significant effects at play.
These might arise from genetic or developmental factors,
or may be governed by the specific functioning of an individual brain.
For instance, pathways between certain motor regions in the brain of a
professional athlete might be exceptionally stronger than that of other
individuals due to prolonged specialized usage of certain circuits. Thus
the deviations might be indicative of plasticity in the
macro-connectome - a hypothesis that requires further exploration. 

In this study we have argued that the inclusion of only the
Poisson-distributed links in a generic representative network for a human
connectome, with the corresponding weights being the link specific Poisson
parameters, is a more meaningful approach than simply obtaining an average
of all structural connectivity matrices. The representative network based
on Poisson parameters at once informs us about the topological
significance, the extent of variability, the generative mechanism and the
functional importance of the underlying links, thus making it far more
useful for further network-theoretic or dynamical analysis. The two
distinct components of the representative network, which we refer as the
basal and the superstructure networks respectively, reveal an altogether
new organizational aspect of the brain. While the source of this dichotomy
within the structural connectome is not clear, one needs to do a more
detailed exploration into the developmental and functional implications of
the two clearly distinguished components that comprise a generic
representative structural connectome of a human brain.


\begin{acknowledgments}
We would like to thank Nitin Williams for his helpful discussions.
SNM has been supported by the IMSc Complex Systems Project ($12^{\rm th}$ 
Plan), and the Center of Excellence in Complex Systems and Data Science, 
both funded by the Department of Atomic Energy, Government of India. The
simulations required for this work were supported by IMSc
High Performance Computing facility (hpc.imsc.res.in) [Nandadevi].
\end{acknowledgments}


\clearpage

\onecolumngrid

\setcounter{figure}{0}
\renewcommand\thefigure{S\arabic{figure}}  
\renewcommand\thetable{S\arabic{table}}

\vspace{1cm}

\begin{center}
\textbf{\large{SUPPLEMENTARY INFORMATION}}
\end{center}

\subsection*{Structure-function correlation for each link}

\begin{figure}[htbp!]
\centering
\centerline{\includegraphics[width=0.8\columnwidth]{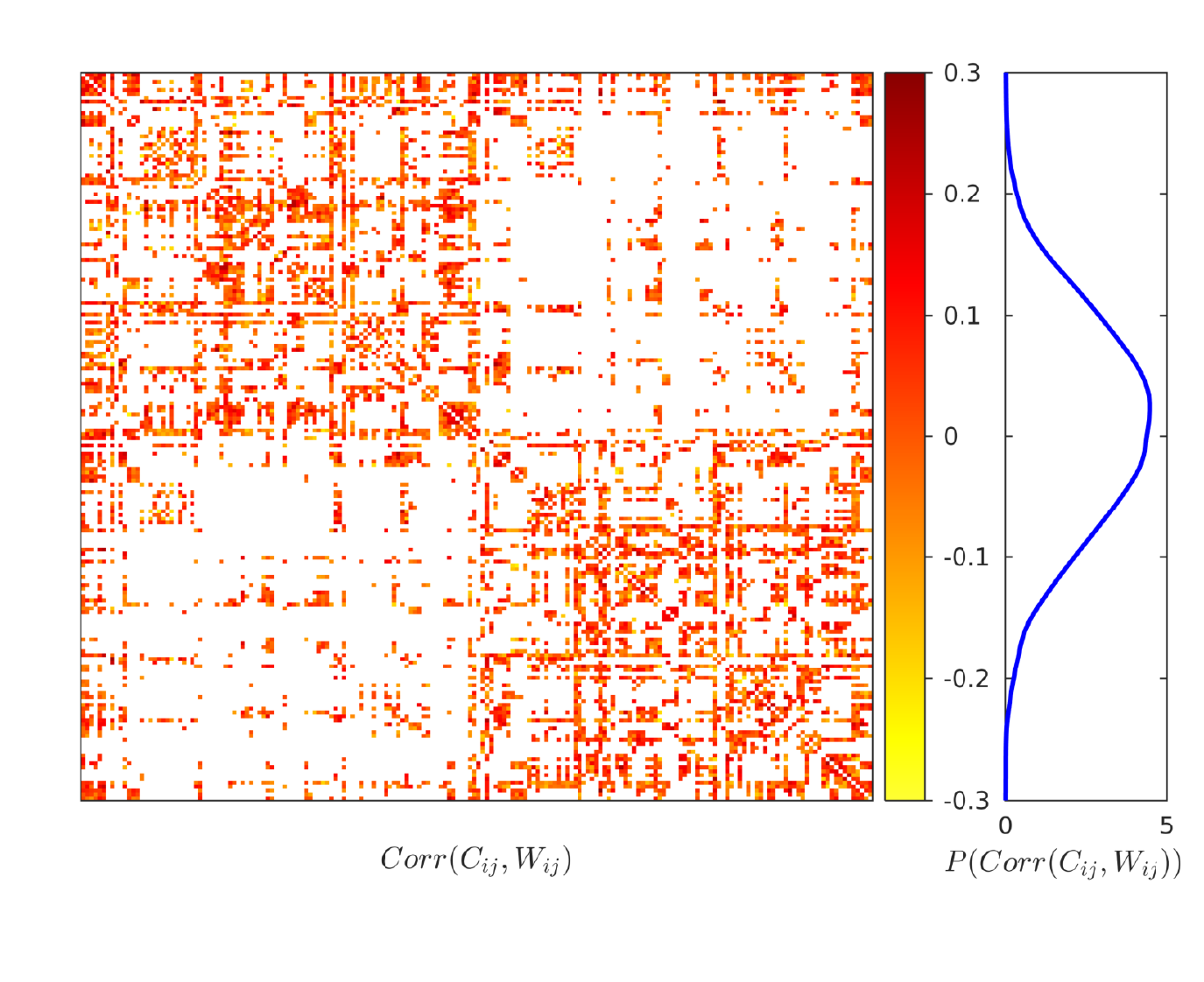}}
\caption[The link-wise correlations between each structural connection and 
its corresponding functional connection over the population are extremely low.]{
\textbf{The link-wise correlations between each structural connection and 
its corresponding functional connection over the population are extremely 
low.}
For each link between
a pair of regions $(i,j)$ we display the correlation between their
structural connection weights $W_{ij}$ and corresponding weights in 
functional connectivities $C_{ij}$, calculated across all individuals (left 
panel). The probability distributions for the correlation values are shown 
in the right panel. Only those links that are identified as part of the 
representative network, i.e., whose weights are Poisson distributed over 
the population, are shown here.   
}
\label{S1}
\end{figure}

In Fig.~\ref{S1}, we display the correspondence between structure and 
function for every link $(i,j)$ belonging to the ``representative'' brain 
network over the population of individuals (see main text). We observe that 
the link-wise correlation $Corr(W_{ij},C_{ij})$ between the structural 
connection weights $W_{ij}$ and corresponding weights in functional 
connectivities $C_{ij}$, calculated across all individuals, tends to be 
extremely low for most of the links. This suggests that the functional 
connectivity has a negligible dependence on the structural connectivity when
observed at the level of nodes and links. This is in contrast with the
macro-level picture shown in Fig. 4 of main text, where we compared the
structural and functional connectivities across entire networks, and found a
comparatively higher and statistically significant correspondence between 
the two.

\subsection*{Community structure in the human structural connectome}

\begin{figure}[bp!]
\centering
\centerline{\includegraphics[width=1\columnwidth]{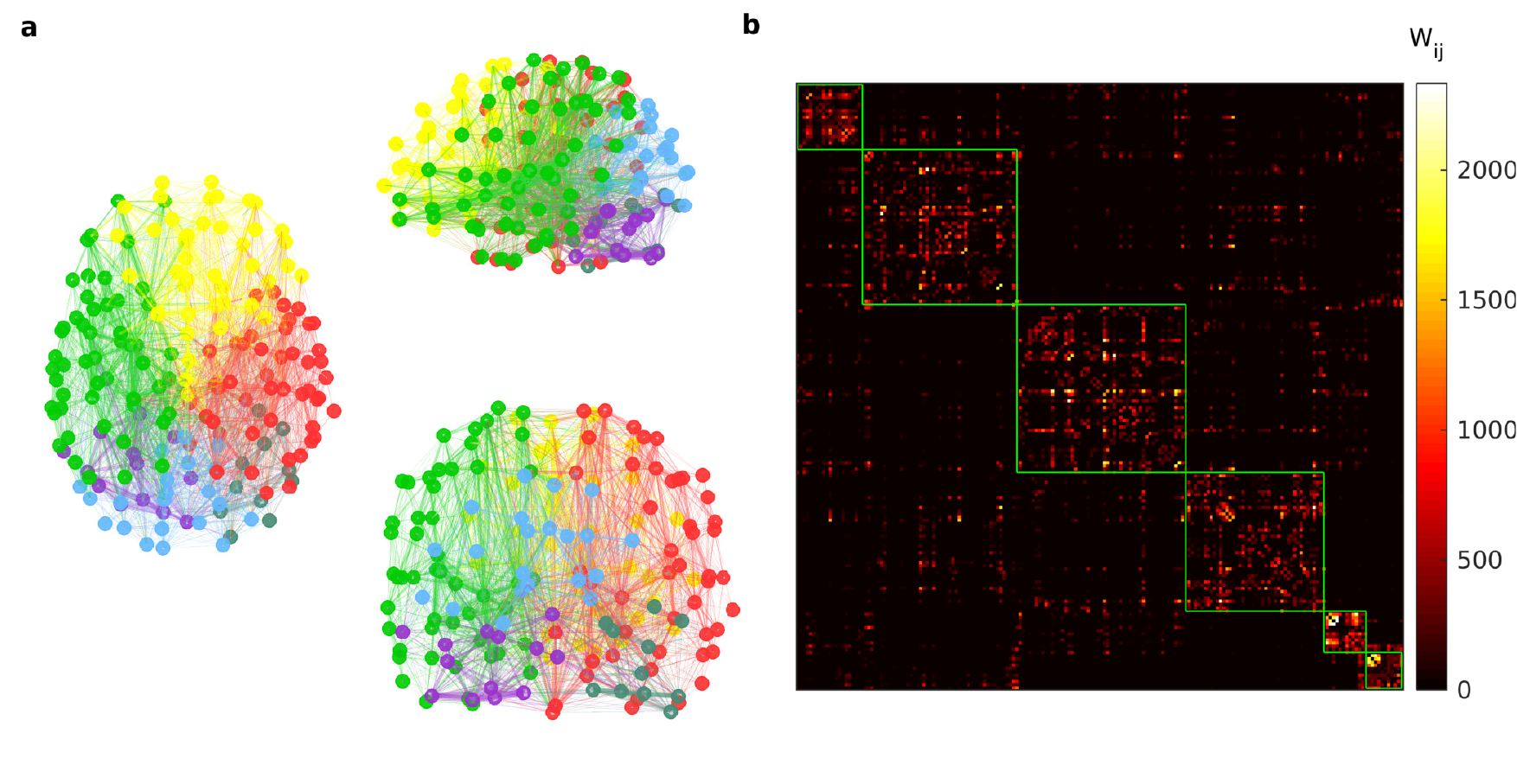}}
\caption[Modules in the structural brain network of an individual subject, 
obtained using Newman Spectral Analysis.]{
\textbf{Modules in the structural brain network of an individual subject, 
obtained using Newman Spectral Analysis.}
(a)~Horizontal, sagittal and coronal projections (left, top right and bottom
right, respectively) of the spatial representations for the structural brain
network of an individual subject, highlighting the $6$ modules obtained from
Newman Spectral Analysis. Here, the nodes are colored in accordance with the
module to which they belong, and the color of each link corresponds to that
of its respective source node, while the thickness of each links is
proportional to its connection weight. Only those links that are part of
the representative network (as described in the main text) are considered here.
(b)~Weighted adjacency matrix representing the network shown in panel~(a).
Here, the nodes are rearranged and grouped according to their modular
membership. The matrix elements are colored in accordance with the
connection weights $W_{ij}$ of the corresponding links (see legend at the
right). The $6$ modules that were obtained from the analysis correspond to
the diagonal blocks, marked by green lines. 
}
\label{S2}
\end{figure}

\begin{figure}[t!]
\centering
\centerline{\includegraphics[width=1\columnwidth]{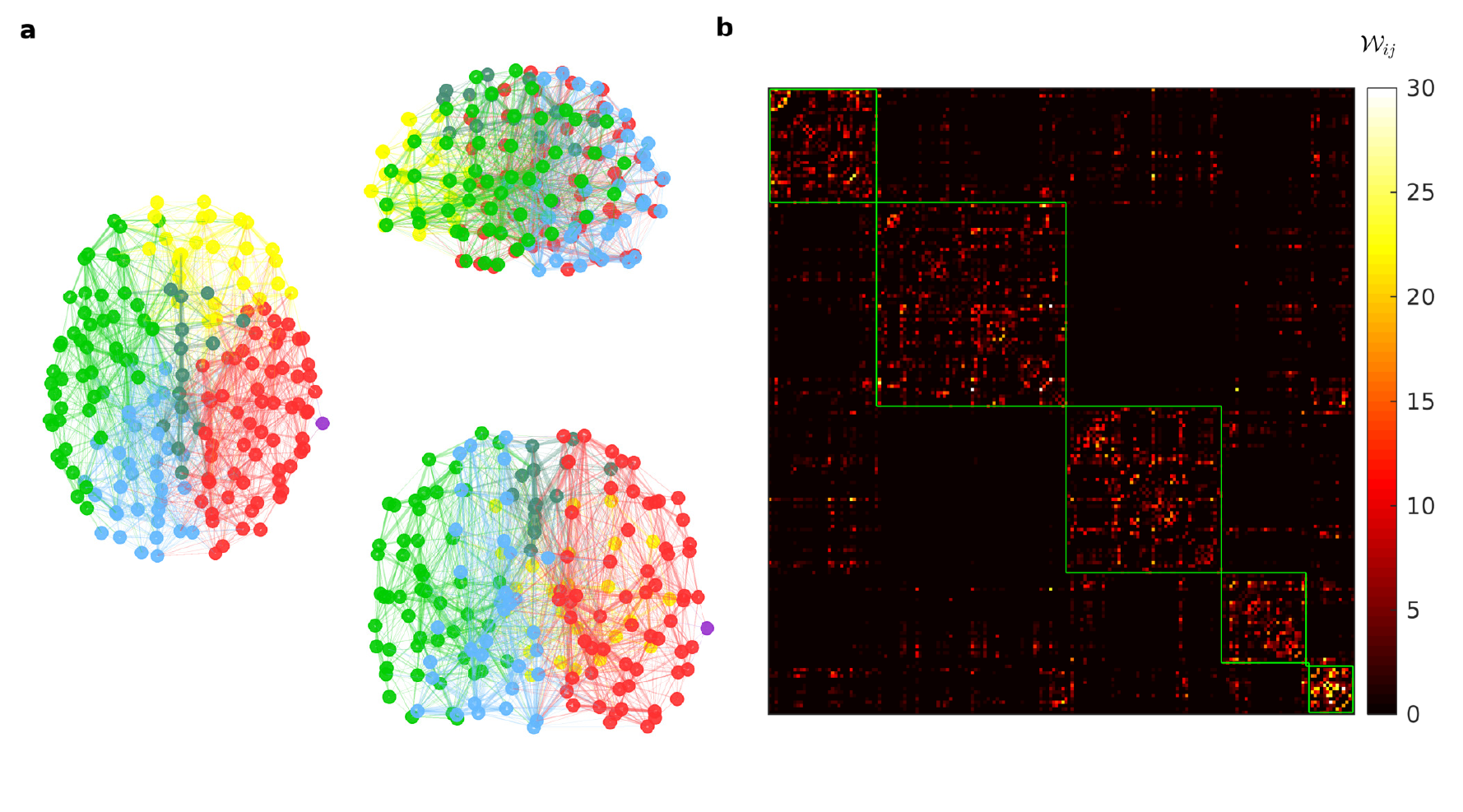}}
\caption[Modules in the rescaled structural brain network of an individual subject, obtained using the Newman Spectral Analysis.]{
\textbf{Modules in the rescaled structural brain network of an individual subject, obtained using the Newman Spectral Analysis.}
(a)~Horizontal, sagittal and coronal projections (left, top, right and 
bottom right, respectively) of the spatial representations for the rescaled 
structural brain network of an individual subject, highlighting the $6$ 
modules obtained from Newman Spectral Analysis. The individual represented 
here is the same as that in Fig.~\ref{S2}. Here, the nodes are colored in 
accordance with the module to which they belong, and the color of each link
corresponds to that of its respective source node, while the thickness of
each links is proportional to its connection weight. Only those links
that are part of the representative network (as described in the main text) are 
considered here.
(b)~Weighted adjacency matrix representing the rescaled network shown in 
panel~(a). Here, the nodes are rearranged and grouped according to their 
modular membership. The matrix elements are colored in accordance with the 
rescaled connection weights $\mathcal{W}_{ij}$ of the corresponding links 
(see legend at the right). The $6$ modules that were obtained from the 
analysis correspond to the diagonal blocks, marked by green lines. 
}
\label{S3}
\end{figure}

We have analyzed two types structural connectivity: the first corresponding
to original connection weights, as given in the database, and the second
consisting of ``rescaled'' connection weights (for details, see Methods). 
We have found modules in the brain network of each individual by
implementing two separate community detection methods which are described 
in~\cite{Pathak2020}, viz., Newman Spectral
Analysis~\cite{Newman2006} and the {\em Infomap Method}~\cite{Rosvall2008}.
We have included only those links that comprise the ``representative''
structural network (as described in the main text). Fig.~\ref{S2} and \ref{S3} 
show the modular decomposition of the original network (Fig.~\ref{S2}) and 
the rescaled network (Fig.~\ref{S3}) for the same individual, as obtained 
from Newman Spectral Analysis.

\begin{figure}[htbp!]
\centering
\centerline{\includegraphics[width=1.2\columnwidth]{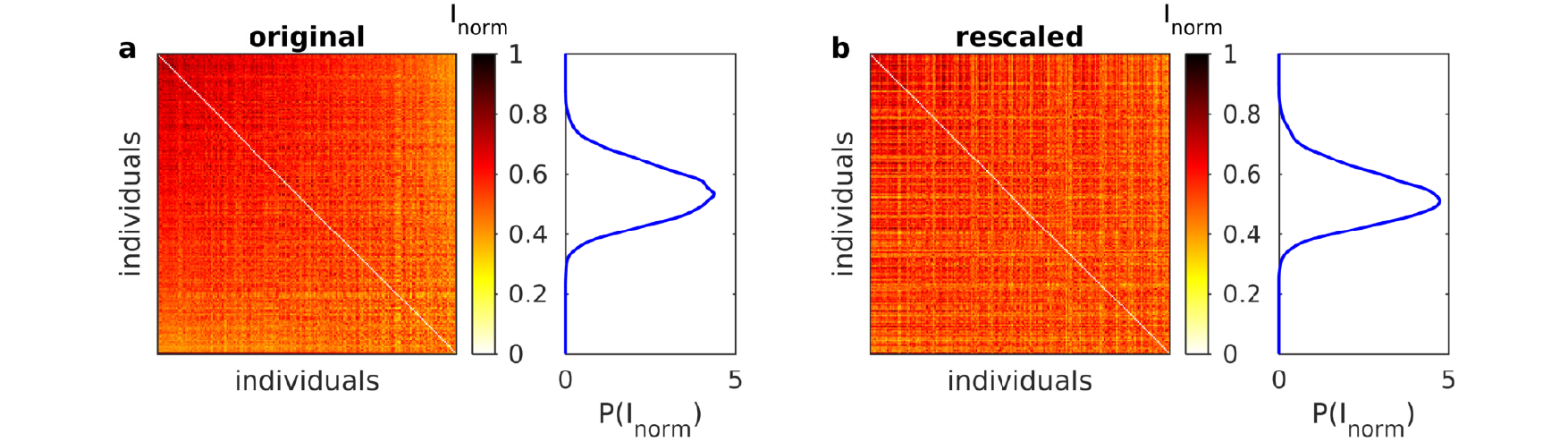}}
\caption[Similarity of Newman Spectral modules across different individual brain networks.]{
\textbf{Similarity of Newman Spectral modules across different individual brain networks.}
Pair-wise values of $I_{norm}$, which quantifies the similarity of modular
paritionings between a pair of individuals, as well as the kernel-smoothened
distributions of the corresponding $I_{norm}$ values, are shown for (a)~the
brain networks having original connection weights, and (b)~brain networks
having rescaled weights. Note that the mode of the $I_{norm}$ distributions
in both cases is $\approx 0.5$, indicating that the  modular decomposition
of individual brain networks varies moderately over the population.
}
\label{S4}
\end{figure}

We observe that the modules obtained are spatially contiguous with clearly defined boundaries. There is only a slight
variation between the modular partitionings of the two types of the networks shown in Fig.~\ref{S2} and \ref{S3}. The similarity of 
modular partitioning between different individuals is shown in  Fig.~\ref{S4} where we show the normalized mutual information $I_{norm}$ 
between all pairs of modular partitionings (for details about normalized mutual information, see~\cite{Pathak2020}).
For most pairs of individuals, the $I_{norm}$ values are $\approx0.5$, which indicates that modular partitioning is 
moderately varying across individuals.


\begin{figure}[htbp!]
\centering
\centerline{\includegraphics[width=1\columnwidth]{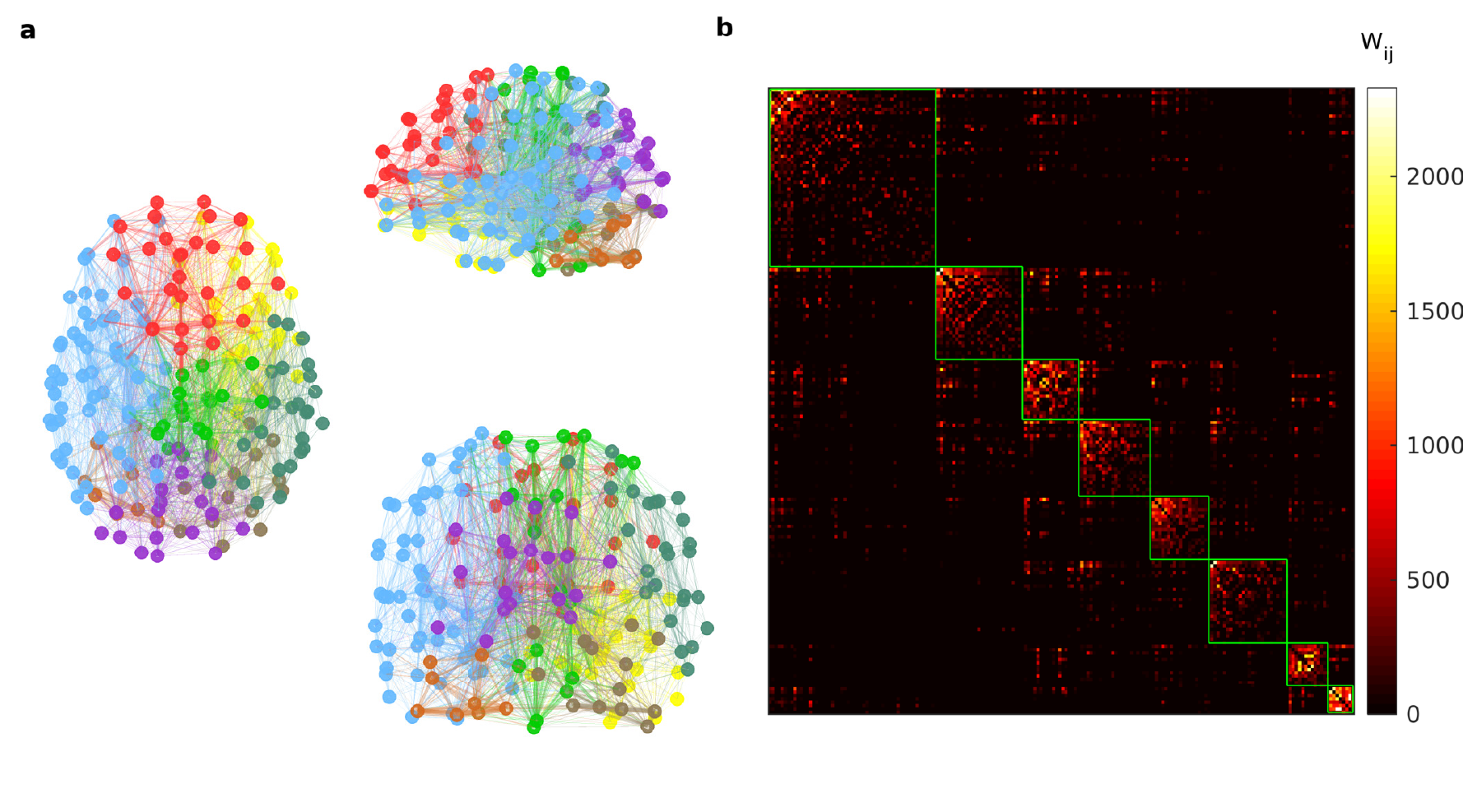}}
\caption[Modules in the structural brain network of an individual subject, obtained using the {\em Infomap} method.]{
\textbf{Modules in the structural brain network of an individual subject, obtained using the {\em Infomap} method.}
(a)~Horizontal, sagittal and coronal projections (left, top, right and 
bottom right, respectively) of the spatial representations for the 
structural brain network of an individual subject, highlighting the $8$ 
modules obtained from the {\em Infomap} method. The individual represented 
here is the same as that in Fig.~\ref{S2}. Here, the nodes are colored in 
accordance with the module to which they belong, and the color of each link 
corresponds to that of its respective source node, while the thickness of
each link is proportional to its connection weight. Only those links 
that are part of the representative network (as described in the main text) are 
considered here.
(b)~Weighted adjacency matrix representing the network shown in panel~(a).
Here, the nodes are rearranged and grouped according to their modular
membership. The matrix elements are colored in accordance with the
connection weights $W_{ij}$ of the corresponding links (see legend at the
right). The $8$ modules that were obtained from the analysis correspond to
the diagonal blocks, marked by green lines. 
}
\label{S5}
\end{figure}

\begin{figure}[htbp!]
\centering
\centerline{\includegraphics[width=1\columnwidth]{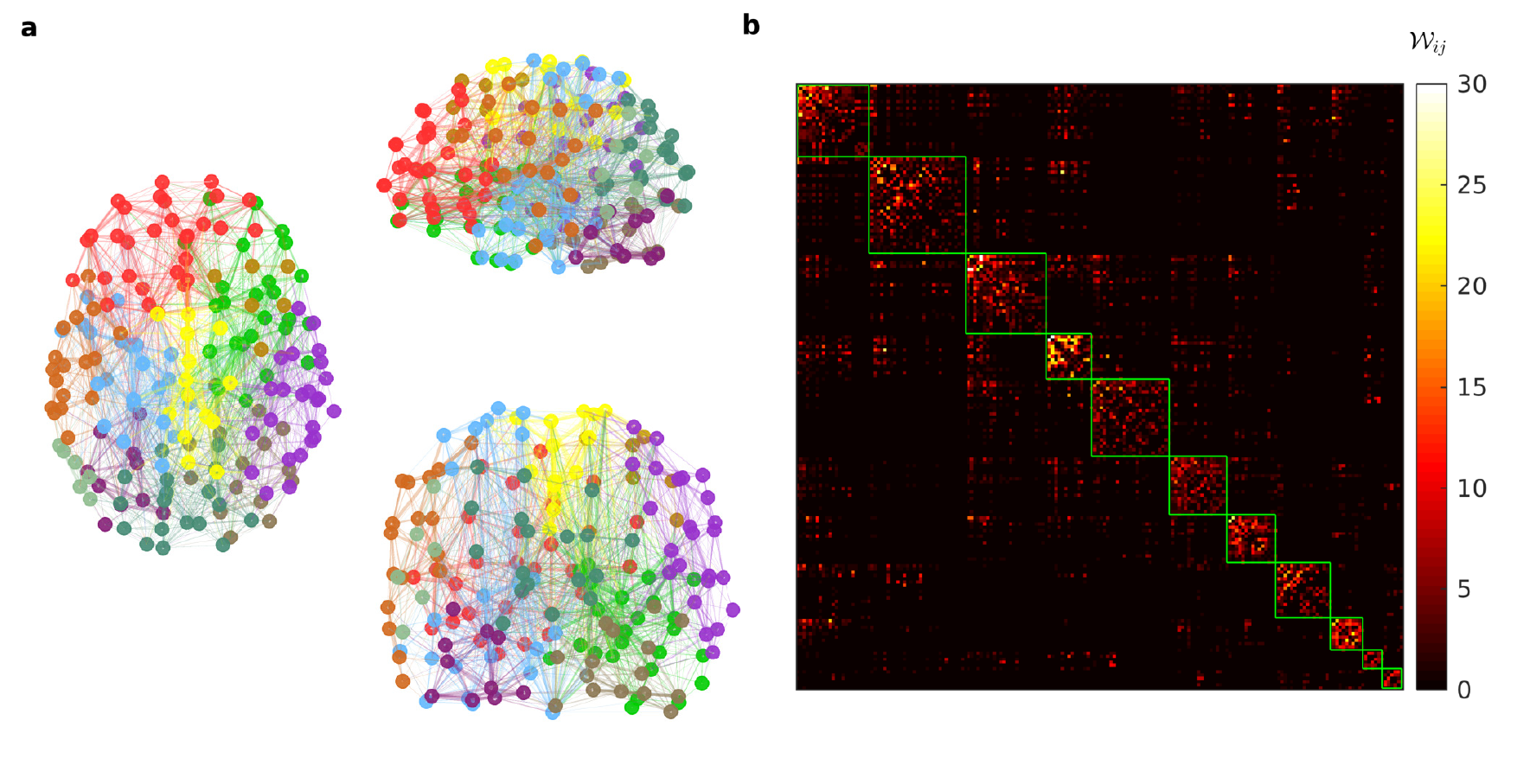}}
\caption[Modules in the rescaled structural brain network of an individual subject, obtained using the {\em Infomap} method.]{
\textbf{Modules in the rescaled structural brain network of an individual subject, obtained using the {\em Infomap} method.}
(a)~Horizontal, sagittal and coronal projections (left, top, right and 
bottom right, respectively) of the spatial representations for the rescaled 
structural brain network of an individual subject, highlighting the $11$ 
modules obtained from the {\em Infomap} method. The individual represented 
here is the same as that in Fig.~\ref{S2}. Here, the nodes are colored in 
accordance with the module to which they belong, and the color of each link
corresponds to that of its respective source node, while the thickness of
each link is proportional to its connection weight. Only those links
that are part of the representative network (as described in the main text) are 
considered here.
(b)~Weighted adjacency matrix representing the rescaled network shown in 
panel~(a). Here, the nodes are rearranged and grouped according to their 
modular membership. The matrix elements are colored in accordance with the 
rescaled connection weights $\mathcal{W}_{ij}$ of the corresponding links 
(see legend at the right). The $11$ modules that were obtained from the 
analysis correspond to the diagonal blocks, marked by green lines. 
}
\label{S6}
\end{figure}

\begin{figure}[htbp!]
\centering
\centerline{\includegraphics[width=1.2\columnwidth]{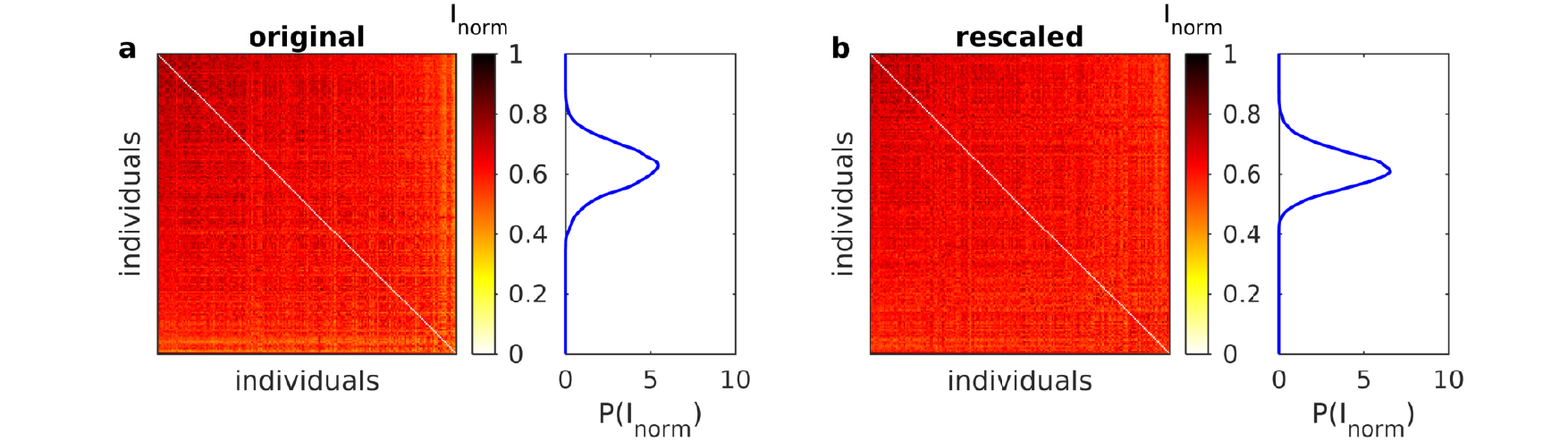}}
\caption[Similarity of {\em Infomap} modules across different individual brain networks.]{
\textbf{Similarity of {\em Infomap} modules across different individual brain networks.}
Pair-wise values of $I_{norm}$, which quantifies the similarity of modular
paritionings between a pair of individuals, as well as the kernel-smoothened
distributions of the corresponding $I_{norm}$ values, are shown for (a)~the
brain networks having original connection weights, and (b)~brain networks
having rescaled weights. Note that the mode of the $I_{norm}$ distributions
in both cases is $\approx 0.6$, indicating that the  modular decomposition
of individual brain networks varies moderately over the population.
}
\label{S7}
\end{figure}

Qualitatively similar results are obtained on using the {\em Infomap} 
method to detect modules in original structural network (Fig.~\ref{S5}) and 
the rescaled structural network (Fig.~\ref{S6}). The networks represented in
Fig.~\ref{S5} and \ref{S6} are from the same individual as that in
Fig.~\ref{S2} and \ref{S3}. Fig.~\ref{S7} shows that the modules obtained
across the individuals using {\em Infomap} method are relatively more
similar to each other, as indicated by higher $I_{norm}$ values, 
in comparison to those obtained using Newman Spectral Analysis. 

\end{document}